\documentclass[10pt,a4paper]{article}

\usepackage[latin1]{inputenc} % permet d'avoir les accents,...
\usepackage[dvips]{graphicx} %permet l'insertion de figures PostScript
\usepackage{a4wide}
\usepackage{latexsym}
\usepackage{amsmath}
\usepackage{amssymb}

\title{Improvement of the $\Theta^+$ width estimation method on the Light Cone}

\author{Cédric Lorcé\\ \small{\emph{Université de
Liège, Institut de Physique, Bât. B5a, B4000 Liège, Belgium}}\\
\small{\emph{Ruhr-Universität Bochum, Institut für Theoretische Physik II, D-44780 Bochum, Germany}}\\
\small{\emph{E-mail: C.Lorce@ulg.ac.be}}}
\date{}

\newcommand{\ud}{\mathrm{d}}

\newcommand{\uM}{\mathcal{M}}
\newcommand{\uZ}{\mathcal{Z}}
\newcommand{\uQcal}{\mathcal{Q}}
\newcommand{\uN}{\mathcal{N}}

\newcommand{\pslash}{p\!\!\!/}

\newcommand{\uPi}{\boldsymbol{\Pi}}

\newcommand{\ux}{\mathbf{x}}

\newcommand{\ur}{\mathbf{r}}
\newcommand{\un}{\mathbf{n}}
\newcommand{\up}{\mathbf{p}}
\newcommand{\uP}{\mathbf{P}}
\newcommand{\uq}{\mathbf{q}}
\newcommand{\uQ}{\mathbf{Q}}

\begin{document}

\maketitle

\begin{center}
\begin{minipage}[t]{15cm}
\small{Recently, Diakonov and Petrov have suggested a formalism in
the Chiral Quark Soliton Model allowing one to derive the 3-, 5-,
7-, \ldots quark wavefunctions for the octet, decuplet and
antidecuplet. They have used this formalism and many strong
approximations in order to estimate the exotic $\Theta^+$ width. The
latter has been estimated to $\sim$4 MeV. Besides they obtained that
the 5-quark component of the nucleon is about 50\% of its 3-quark
component meaning that relativistic effects are not small. We have
improved the technique by taking into account some relativistic
corrections and considering the previously neglected 5-quark
exchange diagrams. We also have computed all nucleon axial charges.
It turns out that exchange diagrams affect very little Diakonov's
and Petrov's results while relativistic corrections reduce the
$\Theta^+$ width to $\sim$2 MeV and the 5- to 3-quark component of
the nucleon ratio to 30\%.}
\end{minipage}
\end{center}

\section{Introduction}

Diakonov and Petrov have derived an effective low-energy Lagrangian
within their instanton model of the QCD vacuum \cite{Action}. This
Lagrangian is presented in Section \ref{Section deux}. It enjoys all
symmetries of chiral QCD, deals with appropriate degrees of freedom
and is believed to reproduce fairly well low-energy QCD physics.
Even though this Lagrangian is a strong simplification of the
original QCD Lagrangian, it is still a considerable task to solve
it. In order to have some insights, the authors have developed a
mean field approach to the problem. A mean field approach is usually
justified by the large number of participants. For example, the
Thomas-Fermi model of atoms is justified at large $Z$
\cite{Thomas-Fermi}. For baryons, the number of colors $N_C$ has
been used as such parameter \cite{Witten}. Since $N_C=3$ in the real
world, one can wonder how accurate is the mean field approach. The
chiral field experiences fluctuations about its mean-field value of
the order of $1/N_C$. These are loop corrections which are further
suppressed by factors of $1/2\pi$ yielding to corrections typically
of the order of 10\%. These are ignored. However, rotations of the
baryon mean field in ordinary and flavor spaces are not small for
$N_C=3$ and are taken into account exactly.

The view that there is a self-consistent mean chiral field in
baryons which binds three constituent quarks \cite{Profile function}
is adopted in this paper. The binding is rather strong: bound-state
quarks are relativistic and their wavefunction has both the upper
$s$-wave Dirac component and the lower $p$-wave Dirac component, see
Section \ref{Section trois}. At the same time, the Dirac sea is
distorted by this mean chiral field leading to the presence of an
indefinite number of additional $q\bar q$ pairs in baryons. Ordinary
baryons are then superpositions of 3-, 5-, 7-, \ldots quark Fock
components. These additional non-perturbative quark-antiquark pairs
are essential for the understanding of the spin crisis and the
nucleon $\sigma$ term \cite{Approximation,Pairs}. The former
experimental value is three times smaller and the latter one is four
times larger than the 3-quark theoretical value \cite{Talk}. This
picture of baryons has been called the Chiral Quark Soliton Model
($\chi$QSM). It leads without any fitting parameters to a reasonable
quantitative description of baryon properties \cite{Profile
function,Baryon properties}, including nucleon parton distributions
at low normalization point \cite{Cutoff} and other baryon
characteristics \cite{Goeke}. The model supports full relativistic
invariance and all symmetries following from QCD.

As mentioned above, the only 3-quark picture of baryons is too
simplistic since it cannot explain some experimental values. It is
then well accepted that one has to consider the effect of additive
quark-antiquark pairs. The problem is now quantitative. A simple
perturbative amount is not sufficient indicating that the
non-perturbative amount is important. $\chi$QSM allows one to
address those questions since it naturally incorporates all additive
quark-antiquark pairs in its description of baryons. On the top of
that, since those additive quark-antiquark pairs are collective
excitations of the mean chiral field, an extra pair costs little
energy. In a recent paper \cite{Original paper} Diakonov and Petrov
have estimated the 5-quark component in the nucleon and found that
it is roughly 50\% of the 3-quark component, and thus not small. The
3-quark picture of the nucleon is then definitely too simplistic.

The quantization of the rotation of the mean chiral field in the
ordinary and flavor spaces yields to correct quantum numbers for the
lowest baryons \cite{Witten}. The rotated mean chiral field can be
represented by $U(\ux)=RV(\ux)R^\dag$ where $R$ is a $SU(3)$
rotation matrix. For simplicity, we deal with $m_s=0$. In this limit
any rotated field is classically as good as the unrotated one. At
the quantum level, the mean chiral field experiences rotations which
cannot be considered as small since there is no cost of energy (zero
modes). These rotations should be quantized properly. As first
pointed out by Witten \cite{Witten} and then derived using different
techniques by a number of authors \cite{Quantization}, the
quantization rule is such that the lowest baryon multiplets are the
octet with spin 1/2 and the decuplet with spin 3/2 followed by the
exotic antidecuplet with spin 1/2. All of those multiplets have same
parity. The lowest baryons are just rotational excitations of the
same mean chiral field (soliton). They are distinguished by their
specific rotational wavefunctions given explicitly in Section
\ref{Rotational section}.

In this approach, most of low-energy properties of the lowest
baryons follow from the shape of the mean chiral field in the
classical baryon. The difference and splitting between baryons are
exclusively due to the difference in their rotational wavefunctions,
difference that can be translated into the quark wavefunctions of
the individual baryons, both in the infinite momentum \cite{Coherent
exponent,Green function} and the rest \cite{Wavefunctions} frames.
In Section \ref{Section trois} we recall the compact general
formalism how to find the 3-, 5-, 7-, \ldots quark wavefunctions
inside the octet, decuplet and antidecuplet baryons and give further
details on the ingredients in Sections \ref{Rotational section},
\ref{Section cinq} and \ref{Section six}. In Section \ref{Section
sept} the 3-quark wavefunctions of the octet and decuplet are shown.
In the non-relativistic limit they are similar to the old $SU(6)$
quark wavefunctions but with well-defined relativistic corrections.
The 5-quark wavefunctions of ordinary and exotic baryons are
presented in Section \ref{section huit}.

We consider baryons in the Infinite Momentum Frame (IMF) since this
is the only frame in which one can distinguish genuine
quark-antiquark pairs of the baryon wavefunctions from vacuum
fluctuations. Therefore an accurate definition of what are the 3-,
5-,7-, \ldots quark Fock components of baryons can be made only in
the IMF. Another advantage of such a frame is that the vector and
axial charges with a finite momentum transfer do not create or
annihilate quarks with infinite momenta. The baryon matrix elements
are thus diagonal in the Fock space.

QCD does not forbid states made of more than 3 quarks as long as
they are colorless. It was first expected that pentaquarks, i.e.
particles which minimal quark content is four quarks and one
antiquark, have wide widths \cite{Wide width1,Wide width2} and then
difficult to observe experimentally. Later, some theorists have
suggested that particular quark structures might exist with a narrow
width \cite{Small width1, Small width2}. The experimental status on
the existence of the exotic $\Theta^+$ pentaquark is still unclear.
There are many experiments in favor (mostly low energy and low
statistics) and against (mostly high energy and high statistics). A
review on the experimental status can be found in
\cite{Experiment1,Experiment2,Experiment3}. Concerning the
experiments in favor, they all agree that the $\Theta^+$ width is
small but give only upper values. It turns out that if it exists,
the exotic $\Theta^+$ has a width of the order of a few MeV or maybe
even less than 1 MeV, a really curious property since usual
resonance widths are of the order of 100 MeV. In the paper
\cite{Small width2} that actually motivated experimentalists to
search a pentaquark, Diakonov, Petrov and Polyakov have estimated
the $\Theta^+$ width to be less than 15 MeV. More recently, Diakonov
and Petrov used the present technique based on light-cone baryon
wavefunction to estimate more accurately the width and have found
that it turns out to be $\sim 4$ MeV \cite{Original paper} and then
the view of a narrow pentaquark resonance within the $\chi$QSM is
safe and appears naturally without any parameter fixing. However,
many approximations have been used such as non-relativistic limit
and omission of some 5-quark contributions (exchange diagrams). The
authors expected that these have high probability to reduce further
the width. This is what has motivated our work. We have improved the
technique in order to include previously neglected diagrams in the
5-quark sector and some relativistic corrections to the
discrete-level wavefunction.

Since the exotic $\Theta^+$ has no 3-quark component and that axial
transitions are diagonal in the Fock space, one has to compute the
5-quark component of the nucleon and the $\Theta^+$. We should add
in principle the contribution coming from the 7-, 9-, \ldots quark
sectors. They are neglected in the present paper. One way to control
the approximation is through the computation of the nucleon axial
charges. The 3-quark values are too crude. The 5-quark contributions
bring the values nearer to experimental ones.

This paper is supposed to be self-consistent. In Sections
\ref{Section neuf} and \ref{Section dix} we remind how to compute
the 3- and 5-quark contributions. We then improve the technique by
taking into account the exchange diagrams and some relativistic
corrections to the discrete-level wavefunction. In section
\ref{Section onze} we collect all old \cite{Original paper} and new
formal results on the strange axial current between the $\Theta^+$
and the nucleon and complete the set of nucleon axial charges. In
Section \ref{Section douze} we give the numerical evaluation of
those observables along with an estimation of the $\Theta^+$ width.
It appears that exchange diagrams, oppositely to what was expected
in \cite{Original paper} have little effect. However relativistic
corrections lead to a reduction of the $\Theta^+$ width to $\sim 2$
MeV and the 5- to 3-quark component of the nucleon ratio to 30\%.

\section{The effective action of the Chiral Quark Soliton
Model}\label{Section deux}

$\chi$QSM is assumed to mimic low-energy QCD thanks to an effective
action describing constituent quarks with a momentum dependent
dynamical mass $M(p)$ interacting with the scalar $\Sigma$ and
pseudoscalar $\uPi$ fields. The chiral circle condition
$\Sigma^2+\uPi^2=1$ is invoked. The momentum dependence of $M(p)$
serves as a formfactor of the constituent quarks and provides also
the effective theory with the UV cutoff. At the same time, it makes
the theory non-local as one can see in the action
\begin{equation}\label{Effective lagrangian}
S_\textrm{eff}=\int\frac{\ud^4p\,\ud^4p'}{(2\pi)^8}\,\bar\psi(p)\left[\pslash\,(2\pi)^4\delta^{(4)}(p-p')-\sqrt{M(p)}(\Sigma(p-p')+i\Pi(p-p')\gamma_5)\sqrt{M(p')}\right]\psi(p')
\end{equation}
where $\psi$ and $\bar\psi$ are quarks fields. This action has been
originally derived in the instanton model of the QCD vacuum
\cite{Action}. Note that oppositely to naive bag picture, this
equation (\ref{Effective lagrangian}) is fully relativistic and
supports all general principles and sum rules for conserved
quantities.

The formfactors $\sqrt{M(p)}$ cut off momenta at some characteristic
scale which corresponds in the instanton picture to the inverse
average size of instantons $1/\bar\rho\approx 600$ MeV. This means
that in the range of quark momenta $p\ll 1/\bar\rho$ one can neglect
the non-locality. We use the standard approach: the constituent
quark mass is replaced by a constant $M=M(0)$ and we mimic the
decreasing function $M(p)$ by the UV Pauli-Villars cutoff
\cite{Cutoff}
\begin{equation}
S_\textrm{eff}=\int\frac{\ud^4p}{(2\pi)^4}\,\bar\psi(p)(\pslash-MU^{\gamma_5})\psi(p)
\end{equation}
with $U^{\gamma_5}$ a $SU(3)$ matrix
\begin{equation}
U^{\gamma_5}=\left(\begin{array}{cc}
                     U_0 & 0 \\
                     0 & 1
                   \end{array}
\right),\qquad U_0=e^{i\pi^a\tau^a\gamma_5}
\end{equation}
and $\tau^a$ being usual $SU(2)$ Pauli matrices.

We are now going to remind the general technique from \cite{Original
paper} that allows one to derive the (ligth-cone) baryon
wavefunctions.

\section{Explicit baryon wavefunction}\label{Section trois}

\begin{figure}[h]\begin{center}\begin{minipage}[c]{8cm}\begin{center}\includegraphics[width=8cm]{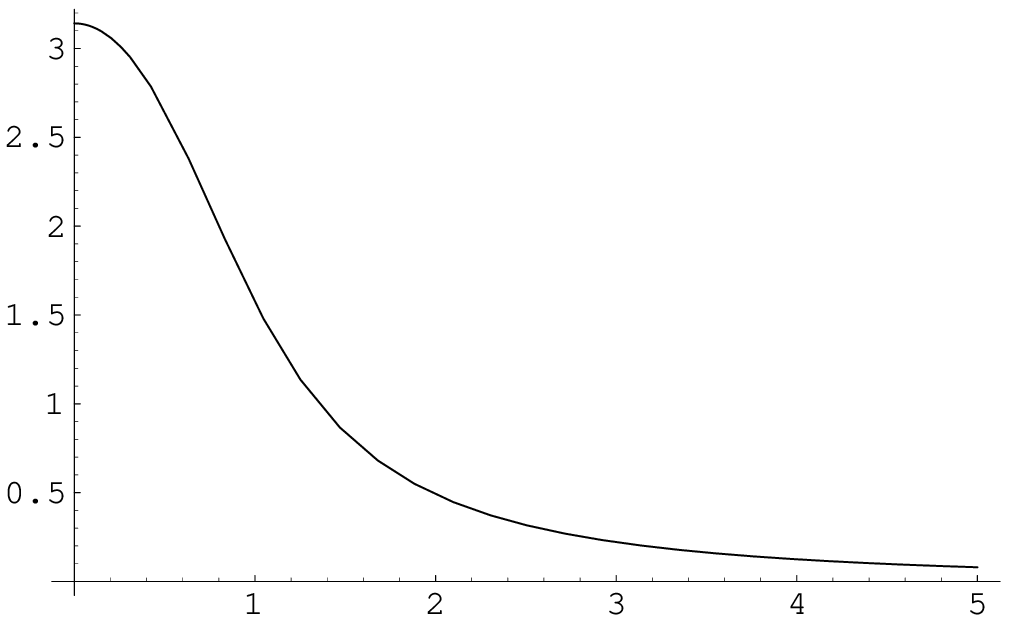}\caption{\small{Profile of the self-consistent
chiral field $P(r)$ in light baryons. The horizontal axis unit is
$r_0=0.8/M=0.46$ fm.\newline\newline\newline}}\label{Profile}
\end{center}\end{minipage}\hspace{0.5cm}
\begin{minipage}[c]{8cm}\begin{center}\includegraphics[width=8cm]{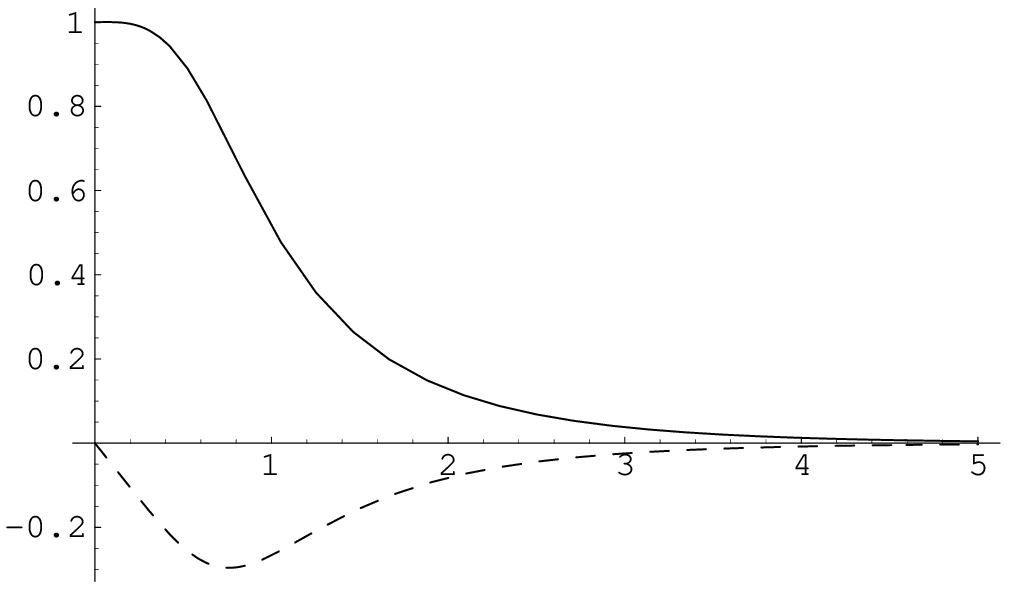}\caption{\small{Upper $s$-wave component $h(r)$ (solid) and lower $p$-wave component
$j(r)$ (dashed) of the bound-state quark level in light baryons.
Each of the three valence quarks has energy $E_\textrm{lev}=200$
MeV. Horizontal axis has units of $1/M=0.57$ fm.}}\label{Level}
\end{center}\end{minipage}\end{center}
\end{figure}

In $\chi$QSM it is easy to define the baryon wavefunction in the
rest frame. Indeed, this model represents quarks in the Hartree
approximation in the self-consistent pion field. The baryon is then
described as $N_C$ valence quarks + Dirac sea in that
self-consistent external field. It has been shown \cite{Coherent
exponent} that the wavefunction of the Dirac sea is the coherent
exponential of the quark-antiquark pairs
\begin{equation}\label{Coherent exponential}
|\Omega\rangle=\exp\left(\int(\ud\up)(\ud\up')\,a^\dag(\up)W(\up,\up')b^\dag(\up')\right)|\Omega_0\rangle
\end{equation}
where $|\Omega_0\rangle$ is the vacuum of quarks and antiquarks
$a,b\,|\Omega_0\rangle=0$, $\langle\Omega_0|\,a^\dag,b^\dag=0$
defined for the quark mass $M\approx 345$ MeV (known to fit numerous
observables within the instanton mechanism of spontaneous chiral
symmetry breaking \cite{Action}), $(\ud\up)=\ud^3\up/(2\pi)^3$ and
$W(\up,\up')$ is the quark Green function at equal times in the
background $\Sigma, \uPi$ fields \cite{Coherent exponent,Green
function} (its explicit expression is given in section \ref{Section
cinq}). In the mean field approximation the chiral field is replaced
by the following spherically-symmetric self-consistent field
\begin{equation}\label{Self-consistent field}
\pi(\ux)=\un\cdot\tau\,P(r),\qquad\un=\ux/r,\qquad
\Sigma(\ux)=\Sigma(r).
\end{equation}
We then have on the chiral circle $\Pi=\un\cdot\tau\sin P(r)$,
$\Sigma(r)=\cos P(r)$ with $P(r)$ being the profile function of the
self-consistent field. The latter is fairly approximated by
\cite{Profile function, Approximation} (see Fig. \ref{Profile})
\begin{equation}\label{Profile function}
P(r)=2\arctan\left(\frac{r_0^2}{r^2}\right),\qquad
r_0\approx\frac{0.8}{M}.
\end{equation}

Such a chiral field creates a bound-state level for quarks, whose
wavefunction $\psi_\textrm{lev}$ satisfies the static Dirac equation
with eigenenergy $E_\textrm{lev}$ in the $K^p=0^+$ sector with
$K=T+J$ \cite{Profile function,Level wavefunction,Valence level}
\begin{equation}
\psi_\textrm{lev}(\ux)=\left(\begin{array}{c}\epsilon^{ji}h(r)\\-i\epsilon^{jk}(\un\cdot\sigma)^i_k\,
j(r)\end{array}\right),\qquad\left\{\begin{array}{c}h'+h\,M\sin
P-j(M\cos P+E_\textrm{lev})=0\\j'+2j/r-j\,M\sin P-h(M\cos
P-E_\textrm{lev})=0\end{array}\right.
\end{equation}
where $i=1,2=\uparrow,\downarrow$ and $j=1,2=u,d$ are respectively
spin and isospin indices. Solving those equations with the
self-consistent field (\ref{Self-consistent field}) one finds that
``valence'' quarks are tightly bound ($E_\textrm{lev}=200$ MeV)
along with a lower component $j(r)$ smaller than the upper one
$h(r)$ (see Fig. \ref{Level}).

For the valence quark part of the baryon wavefunction it suffices to
write the product of $N_C$ quark creation operators that fill in the
discrete level \cite{Coherent exponent}
\begin{equation}\label{Valence part}
\prod_{\textrm{color}=1}^{N_C}\int(\ud\up)F(\up)a^\dag(\up)
\end{equation}
where $F(\up)$ is obtained by expanding and commuting
$\psi_\textrm{lev}(\up)$ with the coherent exponential
(\ref{Coherent exponential})
\begin{equation}\label{Discrete level}
F(\up)=\int(\ud\up')\sqrt{\frac{M}{\epsilon}}[\bar
u(\up)\gamma_0\psi_\textrm{lev}(\up)(2\pi)^3\delta^{(3)}(\up-\up')-W(\up,\up')\bar
v(\up')\gamma_0\psi_\textrm{lev}(-\up')].
\end{equation}
One can see from the second term that the distorted Dirac sea
contributes to the one-quark wavefunction. For the plane-wave Dirac
bispinor $u_\sigma(\up)$ and $v_\sigma(\up)$ we used the standard
basis
\begin{equation}
u_\sigma(\up)=\left(\begin{array}{c}\sqrt{\frac{\epsilon+M}{2M}}s_\sigma\\
\sqrt{\frac{\epsilon-M}{2M}}\frac{\up\cdot\sigma}{|\up|}s_\sigma\end{array}\right),
\qquad
v_\sigma(\up)=\left(\begin{array}{c}\sqrt{\frac{\epsilon-M}{2M}}\frac{\up\cdot\sigma}{|\up|}s_\sigma\\
\sqrt{\frac{\epsilon+M}{2M}}s_\sigma\end{array}\right),\qquad \bar
uu=1=-\bar vv
\end{equation}
where $\epsilon=+\sqrt{\up^2+M^2}$ and $s_\sigma$ are two
2-component spinors normalized to unity
\begin{equation}
s_1=\left(\begin{array}{c}1\\0\end{array}\right),\qquad
s_2=\left(\begin{array}{c}0\\1\end{array}\right).
\end{equation}
The complete baryon wavefunction is then given by the product of the
valence part (\ref{Valence part}) and the coherent exponential
(\ref{Coherent exponential})
\begin{equation}
|\Psi_B\rangle=\prod_{\textrm{color}=1}^{N_C}\int(\ud\up)F(\up)a^\dag(\up)\exp\left(\int(\ud\up)(\ud\up')\,a^\dag(\up)W(\up,\up')b^\dag(\up')\right)|\Omega_0\rangle.
\end{equation}

We remind that the saddle-point of the self-consistent pion field is
degenerate in global translations and global $SU(3)$ flavor
rotations (the $SU(3)$-breaking strange mass can be treated
perturbatively later). These zero modes must be handled with care.
The result is that integrating over translations leads to momentum
conservation which means that the sum of all quarks and antiquarks
momenta have to be equal to the baryon momentum. Integrating over
$SU(3)$ rotations $R$ leads to the projection of the flavor state of
all quarks and antiquarks onto the spin-flavor state $B(R)$ specific
to any particular baryon from the $\left({\bf
8},\frac{1}{2}^+\right)$, $\left({\bf 10},\frac{3}{2}^+\right)$ and
$\left({\bf \overline{10}},\frac{1}{2}^+\right)$ multiplets.

If we restore color ($\alpha=1,2,3$), flavor ($f=1,2,3$), isospin
($j=1,2$) and spin ($\sigma=1,2$) indices, we obtain the following
quark wavefunction of a particular baryon $B$ with spin projection
$k$ \cite{Coherent exponent, Green function}
\begin{eqnarray}
|\Psi_k(B)\rangle&=&\int\ud
R\,B_k^*(R)\epsilon^{\alpha_1\alpha_2\alpha_3}\prod_{n=1}^3\int(\ud\up_n)R_{j_n}^{f_n}F^{j_n\sigma_n}(\up_n)a^\dag_{\alpha_n f_n \sigma_n}(\up_n)\nonumber \\
&\times&\exp\left(\int(\ud\up)(\ud\up')a^\dag_{\alpha f\sigma}(\up)
R^f_j W^{j\sigma}_{j'\sigma'}(\up,\up') R^{\dag
j'}_{f'}b^{\dag\alpha f'\sigma'} \right)|\Omega_0\rangle.\label{Full
glory}
\end{eqnarray}
Then the three $a^\dag$ create three valence quarks with the same
wavefunction $F$ while the $a^\dag$, $b^\dag$ create \emph{any}
number of additional quark-antiquark pairs whose wavefunction is
$W$. One can notice that the valence quarks are antisymmetric in
color whereas additional quark-antiquark pairs are color singlets.
One can obtain the spin-flavor structure of a particular baryon by
projecting a general $qqq+n\,q\bar q$ state onto the quantum numbers
of the baryon under consideration. This projection is an integration
over all spin-flavor rotations $R$ with the rotational wavefunction
$B^*_k(R)$ unique for a given baryon.

Expanding the coherent exponential allows one to get the 3-, 5-, 7-,
\ldots quark wavefunctions of a particular baryon. We still have to
give explicit expressions for the baryon rotational wavefunctions
$B(R)$, the $q\bar q$ pair wavefunction in a baryon
$W^{j\sigma}_{j'\sigma'}(\up,\up')$ and the valence wavefunction
$F^{j\sigma}(\up)$.

\section{Baryon rotational wavefunctions}\label{Rotational section}

Baryon rotational wavefunctions are in general given by the $SU(3)$
Wigner finite-rotation matrices \cite{Wigner matrices} and any
particular projection can be obtained by a $SU(3)$ Clebsch-Gordan
technique. In order to see the symmetries of the quark wavefunctions
explicitly, we keep the expressions for $B(R)$ and integrate over
the Haar measure in eq. (\ref{Full glory}).

The rotational $D$-functions for the $\left({\bf
8},\frac{1}{2}^+\right)$, $\left({\bf 10},\frac{3}{2}^+\right)$ and
$\left({\bf \overline{10}},\frac{1}{2}^+\right)$ multiplets are
listed below in terms of the product of the $R$ matrices. Since the
projection onto a particular baryon in eq. (\ref{Full glory})
involves the conjugate rotational wavefunction, we list the latter
one only. The unconjugate ones are easily obtained by hermitian
conjugation.

\subsection{The octet $\left({\bf
8},\frac{1}{2}^+\right)$}

From the $SU(3)$ group of view, the octet transforms as
$(p,q)=(1,1)$, i.e. the rotational wavefunction can be composed of a
quark (transforming as $R$) and an antiquark (transforming as
$R^\dag$). Then the rotational wavefunction of an octet baryon
having spin index $k=1,2$ is
\begin{equation}
\left[D^{(8,\frac{1}{2})*}(R)\right]^g_{f,k}\sim\epsilon_{kl}R^{\dag
l}_f R^{g}_3.
\end{equation}
The flavor part of this octet tensor $P^g_f$ represents the
particles as follows
\begin{eqnarray}
&P^3_1=N^+_8,\qquad P^3_2=N^0_8,\qquad P^2_1=\Sigma^+_8,\qquad
P^1_2=\Sigma^-_8,&\nonumber\\
&P^1_1=\frac{1}{\sqrt{2}}\,\Sigma^0_8+\frac{1}{\sqrt{6}}\,\Lambda^0_8,\qquad
P^2_2=-\frac{1}{\sqrt{2}}\,\Sigma^0_8+\frac{1}{\sqrt{6}}\,\Lambda^0_8,&\nonumber\\
&P^3_3=-\sqrt{\frac{2}{3}}\,\Lambda^0_8,\qquad P^2_3=\Xi^0_8,\qquad
P^1_3=-\Xi^-_8.&\label{Octet}
\end{eqnarray}
For example, the proton ($f=1,g=3$) and neutron ($f=2,g=3$)
rotational wavefunctions are
\begin{equation}
p_k(R)^*=\sqrt{8}\,\epsilon_{kl}R^{\dag l}_1 R^3_3,\qquad
n_k(R)^*=\sqrt{8}\,\epsilon_{kl}R^{\dag l}_2 R^3_3.
\end{equation}

\subsection{The decuplet $\left({\bf
10},\frac{3}{2}^+\right)$}

The decuplet transforms as $(p,q)=(3,0)$, i.e. the rotational
wavefunction can be composed of three quarks. The rotational
wavefunctions are then labeled by a triple flavor index
$\{f_1f_2f_3\}$ symmetrized in flavor and by a triple spin index
$\{k_1k_2k_3\}$ symmetrized in spin
\begin{equation}
\left[D^{(10,\frac{3}{2})*}(R)\right]_{\{f_1f_2f_3\}\{k_1k_2k_3\}}\sim\epsilon_{k'_1k_1}\epsilon_{k'_2k_2}\epsilon_{k'_3k_3}R^{\dag
k'_1}_{f_1}R^{\dag k'_2}_{f_2}R^{\dag k'_3}_{f_3}\Big|_{\textrm{sym
in }\{f_1f_2f_3\}}.
\end{equation}
The flavor part of this decuplet tensor $D_{f_1f_2f_3}$ represents
the particles as follows
\begin{eqnarray}
&D_{111}=\sqrt{6}\,\Delta^{++}_{10},\qquad
D_{112}=\sqrt{2}\,\Delta^+_{10},\qquad
D_{122}=\sqrt{2}\,\Delta^0_{10},\qquad
D_{222}=\sqrt{6}\,\Delta^-_{10},&\nonumber\\
&D_{113}=\sqrt{2}\,\Sigma^+_{10},\qquad
D_{123}=-\Sigma^0_{10},\qquad
D_{223}=-\sqrt{2}\,\Sigma^-_{10},\qquad
D_{133}=\sqrt{2}\,\Xi^0_{10},&\nonumber\\
&D_{233}=\sqrt{2}\,\Xi^-_{10},\qquad
D_{333}=-\sqrt{6}\,\Omega^-_{10}.&\label{Decuplet}
\end{eqnarray}
For example, the $\Delta^{++}$ with spin projection 3/2
($f_1=1,f_2=1,f_3=1$) and $\Delta^0$ with spin projection 1/2
($f_1=1,f_2=2,f_3=2$) rotational wavefunctions are
\begin{equation}
\Delta^{++}_{\uparrow\uparrow\uparrow}(R)^*=\sqrt{10}\,R^{\dag
2}_1R^{\dag 2}_1R^{\dag 2}_1,\qquad
\Delta^0_\uparrow(R)^*=\sqrt{10}\,R^{\dag 2}_2(2R^{\dag 2}_1R^{\dag
1}_2+R^{\dag 2}_2R^{\dag 1}_1).
\end{equation}

\subsection{The antidecuplet $\left({\bf\overline{
10}},\frac{1}{2}^+\right)$}

The antidecuplet transforms as $(p,q)=(0,3)$, i.e. the rotational
wavefunction can be composed of three antiquarks. The rotational
wavefunctions are then labeled by a triple flavor index
$\{f_1f_2f_3\}$ symmetrized in flavor
\begin{equation}
\left[D^{(\overline{10},\frac{1}{2})*}(R)\right]^{\{f_1f_2f_3\}}_k\sim
R^{f_1}_3R^{f_2}_3R^{f_3}_k\Big|_{\textrm{sym in }\{f_1f_2f_3\}}.
\end{equation}
The flavor part of this antidecuplet tensor $T^{f_1f_2f_3}$
represents the particles as follows
\begin{eqnarray}
&T^{111}=\sqrt{6}\,\Xi^{--}_{\overline{10}},\qquad
T^{112}=-\sqrt{2}\,\Xi^-_{\overline{10}},\qquad
T^{122}=\sqrt{2}\,\Xi^0_{\overline{10}},\qquad
T^{222}=-\sqrt{6}\,\Xi^+_{\overline{10}},&\nonumber\\
&T^{113}=\sqrt{2}\,\Sigma^-_{\overline{10}},\qquad
T^{123}=-\Sigma^0_{\overline{10}},\qquad
T^{223}=-\sqrt{2}\,\Sigma^+_{\overline{10}},\qquad
T^{133}=\sqrt{2}\,N^0_{\overline{10}},&\nonumber\\
&T^{233}=-\sqrt{2}\,N^+_{\overline{10}},\qquad
T^{333}=\sqrt{6}\,\Theta^+_{\overline{10}}.&\label{Antidecuplet}
\end{eqnarray}
For example, the $\Theta^+$ ($f_1=3,f_2=3,f_3=3$) and neutron$^*$
from $\overline{10}$ ($f_1=1,f_2=3,f_3=3$) rotational wavefunctions
are
\begin{equation}
\Theta^+_k(R)^*=\sqrt{30}\,R^3_3R^3_3R^3_k,\qquad
n^{\overline{10}}_k(R)^*=\sqrt{10}\,R^3_3(2R^1_3R^3_k+R^3_3R^1_k).
\end{equation}

All examples of rotational wavefunctions above have been normalized
in such a way that for any (but the same) spin projection we have
\begin{equation}
\int\ud R\,B^*_\textrm{spin}(R)B^\textrm{spin}(R)=1,
\end{equation}
the integral being zero for different spin projections. Note that
rotational wavefunctions belonging to different baryons are also
orthogonal. This can be easily checked using the group integrals in
Appendix A. The particle representations (\ref{Octet}),
(\ref{Decuplet}) and (\ref{Antidecuplet}) were found in \cite{Tensor
structure}.

\section{$q\bar q$ pair wavefunction}\label{Section cinq}

In \cite{Coherent exponent, Green function} it is explained that the
pair wavefunction $W^{j\sigma}_{j'\sigma'}(\up,\up')$ is expressed
by means of the finite-time quark Green function at equal times in
the external static chiral field (\ref{Self-consistent field}). The
Fourier transforms of this field will be needed
\begin{equation}\label{Fourier tranform of mean field}
\Pi(\uq)^j_{j'}=\int\ud^3\ux\,
e^{-i\uq\cdot\ux}(\un\cdot\tau)^j_{j'}\sin
P(r),\qquad\Sigma(\uq)^j_{j'}=\int\ud^3\ux\, e^{-i\uq\cdot\ux}(\cos
P(r)-1)\delta^j_{j'}
\end{equation}
where $\Pi(\uq)$ is purely imaginary and odd and $\Sigma(\uq)$ is
real and even.

A simplified interpolating approximation for the pair wavefunction
$W$ has been derived in \cite{Coherent exponent, Green function} and
becomes exact in three limiting cases: i) small pion field $P(r)$,
ii) slowly varying $P(r)$ and iii) fast varying $P(r)$. Since the
model is relativistically invariant, this wavefunction can be
translated to the infinite momentum frame (IMF). In this particular
frame, the result is a function of the fractions of the baryon
longitudinal momentum carried by the quark $z$ and antiquark $z'$ of
the pair and their transverse momenta $\up_\perp$, $\up'_\perp$
\begin{equation}\label{Pair wavefunction}
W^{j,\sigma}_{j'\sigma'}(z,\up_\perp;z',\up'_\perp)=\frac{M\uM}{2\pi
Z}\left\{\Sigma^j_{j'}(\uq)[M(z'-z)\tau_3+\uQ_\perp\cdot\tau_\perp]^\sigma_{\sigma'}+i\Pi^j_{j'}(\uq)[-M(z'+z)\bold
1+i\uQ_\perp\times\tau_\perp]^\sigma_{\sigma'}\right\}
\end{equation}
where $\uq=((\up+\up')_\perp,(z+z')\uM)$ is the three-momentum of
the pair as a whole transferred from the background fields
$\Sigma(\uq)$ and $\Pi(\uq)$, $\tau_{1,2,3}$ are Pauli matrices,
$\uM$ is the baryon mass and $M$ is the constituent quark mass. In
order to condense the notations we used
\begin{equation}\label{Notation}
Z=\uM^2zz'(z+z')+z(p'^2_\perp+M^2)+z'(p^2_\perp+M^2),\qquad
\uQ_\perp=z\up'_\perp-z'\up_\perp.
\end{equation}

This pair wavefunction $W$ is normalized in such a way that the
creation-annihilation operators satisfy the following
anticommutation relations
\begin{equation}\label{Anticommutation}
\{a^{\alpha_1f_1\sigma_1}(z_1,\up_{1\perp}),a^\dag_{\alpha_2f_2\sigma_2}(z_2,\up_{2\perp})\}=\delta^{\alpha_1}_{\alpha_2}\delta^{f_1}_{f_2}\delta^{\sigma_1}_{\sigma_2}
\delta(z_1-z_2)(2\pi)^2\delta^{(2)}(\up_{1\perp}-\up_{2\perp})
\end{equation}
and similarly for $b$, $b^\dag$, the integrals over momenta being
understood as $\int\ud z\int\ud^2\up_\perp/(2\pi)^2$.

\section{Discrete-level wavefunction}\label{Section six}

We see from eq. (\ref{Discrete level}) that the discrete-level
wavefunction
$F^{j\sigma}(\up)=F^{j\sigma}_\textrm{lev}(\up)+F^{j\sigma}_\textrm{sea}(\up)$
is the sum of two parts: the one is directly the wavefunction of the
valence level and the other is related to the change of the number
of quarks at the discrete level due to the presence of the Dirac
sea; it is a relativistic effect and can be ignored in the
non-relativistic limit ($E_\textrm{lev}\approx M$) together with the
small $L=1$ lower component $j(r)$. Indeed, in the baryon rest frame
$F^{j\sigma}_\textrm{lev}$ gives
\begin{equation}
F^{j\sigma}_\textrm{lev}=\epsilon^{j\sigma}\left(\sqrt{\frac{E_\textrm{lev}+M}{2E_\textrm{lev}}}h(p)+\sqrt{\frac{E_\textrm{lev}-M}{2E_\textrm{lev}}}j(p)\right)
\end{equation}
where $h(p)$ and $j(p)$ are the Fourier transforms of the valence
wavefunction
\begin{eqnarray}
h(p)&=&\int\ud^3\ux\, e^{-i\up\cdot\ux}h(r)=4\pi\int_0^\infty\ud
r\,r^2\frac{\sin pr}{pr}h(r),\\
j^a(p)&=&\int\ud^3\ux\,
e^{-i\up\cdot\ux}(-in^a)j(r)=\frac{p^a}{|\up|}j(p),\qquad
j(p)=\frac{4\pi}{p^2}\int_0^\infty\ud r\,(pr\cos pr-\sin pr)j(r).
\end{eqnarray}
In the non-relativistic limit the second term is double-suppressed:
first due to the kinematical factor and second due to the smallness
of the $L=1$ wave $j(r)$ compared to the $L=0$ wave $h(r)$.

Switching to the IMF one obtains \cite{Coherent exponent, Green
function}
\begin{equation}\label{Discrete level IMF}
F^{j\sigma}_\textrm{lev}(z,\up_\perp)=\sqrt{\frac{\uM}{2\pi}}\left[\epsilon^{j\sigma}h(p)+(p_z\bold
1+i\up_\perp\times\tau_\perp)^\sigma_{\sigma'}\epsilon^{j\sigma'}\frac{j(p)}{|\up|}\right]_{p_z=z\uM-E_\textrm{lev}}.
\end{equation}

The ``sea'' part of the discrete-level wavefunction gives in the IMF
\begin{equation}\label{Discrete level IMF 2}
F^{j\sigma}_\textrm{sea}(z,\up_\perp)=-\sqrt{\frac{\uM}{2\pi}}\int\ud
z'\frac{\ud^2\up'_\perp}{(2\pi)^2}\,W^{j\sigma}_{j'\sigma'}(z,\up_\perp;z',\up'_\perp)\,\epsilon^{j'\sigma''}\left[(\tau_3)^{\sigma'}_{\sigma''}h(p')-(\up'\cdot\tau)^{\sigma'}_{\sigma''}\frac{j(p')}{|\up'|}\right]_{p_z=z\uM-E_\textrm{lev}}.
\end{equation}

In the work made by Diakonov and Petrov \cite{Original paper}, the
relativistic effects in the discrete-level wavefunction were
neglected. One can then use only the first term in (\ref{Discrete
level IMF})
\begin{equation}\label{Approximation}
F^{j\sigma}(z,\up_\perp)\approx\sqrt{\frac{\uM}{2\pi}}\,\epsilon^{j\sigma}h(p)\big|_{p_z=z\uM-E_\textrm{lev}}.
\end{equation}

\section{3-quark components of baryons}\label{Section sept}

It will be shown in this section how to derive systematically the
3-quark component of the octet and decuplet baryons (antidecuplet
baryons have no such component) and that they become in the
non-relativistic limit similar to the well-known $SU(6)$
wavefunctions of the constituent quark model.

\begin{figure}[h]\begin{center}\begin{minipage}[c]{8cm}\begin{center}\includegraphics[width=2.5cm]{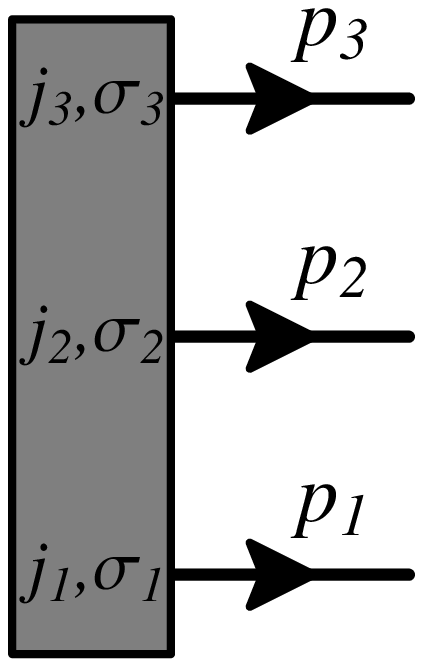}
\caption{\small{Schematic representation of the 3-quark component of
baryon wavefunctions. The dark gray rectangle stands for the three
discrete-level wavefunctions
$F^{j_i\sigma_i}(\up_i)$.}}\label{Threequarks}
\end{center}\end{minipage}\hspace{0.5cm}
\begin{minipage}[c]{8cm}\begin{center}\includegraphics[width=2cm]{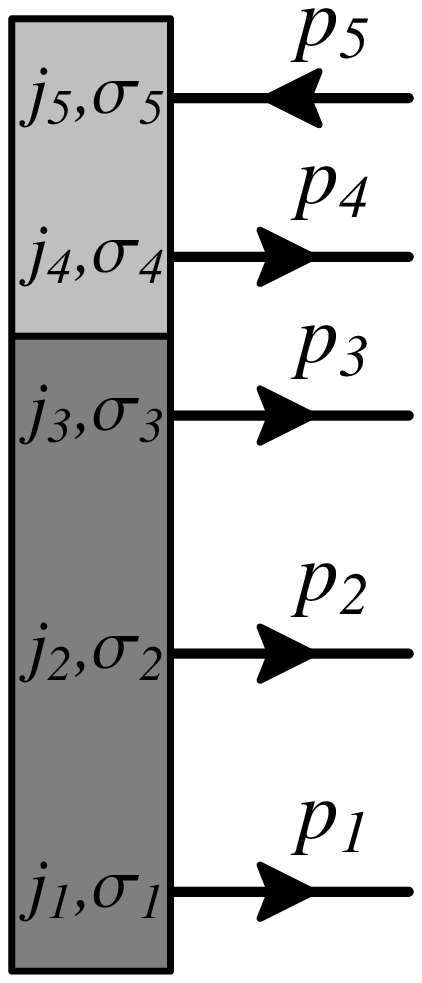}
\caption{\small{Schematic representation of the 5-quark component of
baryon wavefunctions. The light gray rectangle stands for the pair
wavefunction $W^{j_i\sigma_i}_{j_k\sigma_k}(\up_i,\up_k)$ where the
reversed arrow represents the antiquark.}}\label{Fivequarks}
\end{center}\end{minipage}\end{center}
\end{figure}

An expansion of the coherent exponential (\ref{Coherent
exponential}) gives access to all Fock components of the baryon
wavefunction. Since we are interested in the present case only in
the 3-quark component, this coherent exponential is just ignored.
One can see from eq. (\ref{Full glory}) that the three valence
quarks are rotated by the $SU(3)$ matrices $R^f_j$ where
$f=1,2,3=u,d,s$ is the flavor and $j=1,2=u,d$ is the isospin index.
The projection onto a specific baryon leads to the following group
integral
\begin{equation}\label{T tensor}
T(B)^{f_1f_2f_3}_{j_1j_2j_3,k}\equiv\int\ud
R\,B^*_k(R)R^{f_1}_{j_1}R^{f_2}_{j_2}R^{f_3}_{j_3}.
\end{equation}
The group integrals can be found in Appendix A. This tensor $T$ must
be contracted with the three discrete-level wavefunctions
\begin{equation}
F^{j_1\sigma_1}(\up_1)F^{j_2\sigma_2}(\up_2)F^{j_3\sigma_3}(\up_3).
\end{equation}
The wavefunction is schematically represented on Fig.
\ref{Threequarks}.

For example, one obtains the following non-relativistic 3-quark
wavefunction for the neutron in the coordinate space
\begin{equation}
(|n\rangle_k)^{f_1f_2f_3,\sigma_1\sigma_2\sigma_3}(\ur_1,\ur_2,\ur_3)=\frac{\sqrt{8}}{24}\,\epsilon^{f_1f_2}\epsilon^{\sigma_1\sigma_2}\delta^{f_3}_2\delta^{\sigma_3}_kh(r_1)h(r_2)h(r_3)+\textrm{
permutations of 1,2,3}
\end{equation}
times the antisymmetric tensor $\epsilon^{\alpha_1\alpha_2\alpha_3}$
in color. This equation says that in the 3-quark picture the whole
neutron spin is carried by a $d$-quark while the $ud$ pair is in the
spin- and isopin-zero combination. This is similar to the better
known non-relativistic $SU(6)$ wavefunction of the neutron
\begin{eqnarray}
|n\uparrow\rangle&=&2 |d\uparrow(r_1)\rangle |d\uparrow(r_2)\rangle
|u\downarrow(r_3)\rangle-|d\uparrow(r_1)\rangle
|u\uparrow(r_2)\rangle |d\downarrow(r_3)\rangle-
|u\uparrow(r_1)\rangle |d\uparrow(r_2)\rangle
|d\downarrow(r_3)\rangle\nonumber\\&+&\textrm{ permutations of
1,2,3}.
\end{eqnarray}
There are, of course, many relativistic corrections arising from the
exact discrete-level wavefunction (\ref{Discrete level
IMF},\ref{Discrete level IMF 2}) and the additional quark-antiquark
pairs, both effects being generally not small.

\section{5-quark components of baryons}\label{section huit}

The 5-quark component of the baryon wavefunctions is obtained by
expanding the coherent exponential (\ref{Coherent exponential}) to
the linear order in the $q\bar q$ pair. The projection involves now
along with the three $R$'s from the discrete level two additional
matrices $R\,R^\dag$ that rotate the quark-antiquark pair in the
$SU(3)$ space
\begin{equation}
T(B)^{f_1f_2f_3f_4,j_5}_{j_1j_2j_3j_4,f_5,k}\equiv\int\ud
R\,B^*_k(R)R^{f_1}_{j_1}R^{f_2}_{j_2}R^{f_3}_{j_3}R^{f_4}_{j_4}R^{\dag
j_5}_{f_5}.
\end{equation}
One then obtain the following 5-quark component of the neutron
wavefunction in the momentum space
\begin{eqnarray}
(|n\rangle_k)^{f_1f_2f_3f_4,\sigma_1\sigma_2\sigma_3\sigma_4}_{f_5,\sigma_5}(\up_1\ldots\up_5)&=&\frac{\sqrt{8}}{360}\,F^{j_1\sigma_1}(\up_1)F^{j_2\sigma_2}(\up_2)F^{j_3\sigma_3}(\up_3)W^{j_4\sigma_4}_{j_5\sigma_5}(\up_4,\up_5)\nonumber\\
&\times&\epsilon_{k'k}\left\{\epsilon^{f_1f_2}\epsilon_{j_1j_2}\left[\delta^{f_3}_2\delta^{f_4}_{f_5}\left(4\delta^{j_5}_{j_4}\delta^{k'}_{j_3}-\delta^{j_5}_{j_3}\delta^{k'}_{j_4}\right)+\delta^{f_4}_2\delta^{f_3}_{f_5}\left(4\delta^{j_5}_{j_3}\delta^{k'}_{j_4}-\delta^{j_5}_{j_4}\delta^{k'}_{j_3}\right)\right]\right.\nonumber\\
&+&\left.\epsilon^{f_1f_4}\epsilon_{j_1j_4}\left[\delta^{f_2}_2\delta^{f_3}_{f_5}\left(4\delta^{j_5}_{j_3}\delta^{k'}_{j_2}-\delta^{j_5}_{j_2}\delta^{k'}_{j_3}\right)+\delta^{f_3}_2\delta^{f_2}_{f_5}\left(4\delta^{j_5}_{j_2}\delta^{k'}_{j_3}-\delta^{j_5}_{j_3}\delta^{k'}_{j_2}\right)\right]\right\}\nonumber\\
&+&\textrm{permutations of 1,2,3}.
\end{eqnarray}
The color degrees of freedom are not explicitly written but the
three valence quarks (1,2,3) are still antisymmetric in color while
the quark-antiquark pair (4,5) is a color singlet. The wavefunction
is schematically represented on Fig. \ref{Fivequarks}.

Exotic baryons from the
$\left({\bf\overline{10}},\frac{1}{2}^+\right)$ multiplet, despite
the inexistence of a 3-quark component, have such a 5-quark
component in their wavefunction. One has for example the following
wavefunction for the $\Theta^+$
\begin{eqnarray}
(|\Theta^+\rangle_k)^{f_1f_2f_3f_4,\sigma_1\sigma_2\sigma_3\sigma_4}_{f_5,\sigma_5}(\up_1\ldots\up_5)&=&\frac{\sqrt{30}}{180}\,F^{j_1\sigma_1}(\up_1)F^{j_2\sigma_2}(\up_2)F^{j_3\sigma_3}(\up_3)W^{j_4\sigma_4}_{j_5\sigma_5}(\up_4,\up_5)\nonumber\\
&\times&\epsilon^{f_1f_2}\epsilon^{f_3f_4}\epsilon_{j_1j_2}\epsilon_{j_3j_4}\delta^3_{f_5}\delta^{j_5}_k\nonumber\\
&+&\textrm{ permutations of 1,2,3}.
\end{eqnarray}
The color structure is here very simple:
$\epsilon^{\alpha_1\alpha_2\alpha_3}\delta^{\alpha_4}_{\alpha_5}$.
This wavefunction says that we have two $ud$ pairs in the spin- and
isospin-zero combination and that the whole $\Theta^+$ spin is
carried by the $\bar s$ quark. One has naturally obtained the
minimal quark content of the $\Theta^+$ pentaquark $uudd\bar s$.

\section{Normalizations, vector and axial charges}\label{Section
neuf}

The normalization of a Fock component $n$ of a specific
spin-$\frac{1}{2}$ baryon $B$ wavefunction is obtained by
\begin{equation}\label{Normalization}
\uN^{(n)}(B)=\frac{1}{2}\,\delta^k_l\langle\Psi^{(n)l}(B)|\Psi^{(n)}_k(B)\rangle.
\end{equation}
One has to drag all annihilation operators in $\Psi^{(n)\dag l}(B)$
to the right and the creation operators in $\Psi^{(n)}_k(B)$ to the
left so that the vacuum state $|\Omega_0\rangle$ is nullified. One
then gets a non-zero result due to the anticommutation relations
(\ref{Anticommutation}) or equivalently to the ``contractions'' of
the operators.

A typical physical observable is the matrix element of some operator
(preferably written in terms of quark annihilation-creation
operators $a$, $b$, $a^\dag$, $b^\dag$) sandwiched between the
initial and final baryon wavefunctions. As Diakonov and Petrov did
in their paper \cite{Original paper}, we shall consider only the
operators of the vector and axial charges which can be written as
\begin{eqnarray}
\left\{\begin{array}{c}Q\\Q_5\end{array}\right\}=\int\ud^3\ux\,\bar\psi_eJ^e_h\left\{\begin{array}{c}\gamma_0\\
\gamma_0\gamma_5\end{array}\right\}\psi^h &=&\int\ud
z\,\frac{\ud^2\up_\perp}{(2\pi)^2}\left[a^\dag_{e\pi}(z,\up_\perp)a^{h\rho}(z,\up_\perp)J^e_h\left\{\begin{array}{c}\delta^\pi_\rho\\(-\sigma_3)^\pi_\rho\end{array}\right\}
\right.\nonumber\\&-&\left.b^{\dag
h\rho}(z,\up_\perp)b_{e\pi}(z,\up_\perp)J^e_h\left\{\begin{array}{c}\delta^\pi_\rho\\(-\sigma_3)^\pi_\rho\end{array}\right\}\right]
\end{eqnarray}
where $J^e_h$ is the flavor content of the charge and
$\pi,\rho=1,2=L,R$ are helicity states. Notice that there are
neither $a^\dag b^\dag$ nor $ab$ terms in the charges. This is a
great advantage of the IMF where the number of $q\bar q$ pairs is
not changed by the current. Hence there will only be diagonal
transitions in the Fock space, i.e. the charges can be decomposed
into the sum of the contributions from all Fock components
$Q=\sum_nQ^{(n)}$, $Q_5=\sum_nQ^{(n)}_5$. Notice that there is also
a color index which is just summed up.

The axial charges of the nucleon are defined as forward matrix
elements of the axial current
\begin{equation}
\langle
N(p)|\bar\psi\gamma_\mu\gamma_5\lambda^a\psi|N(p)\rangle=g^{(a)}_A\bar
u(p)\gamma_\mu\gamma_5 u(p)
\end{equation}
where $a=0,3,8$ and $\lambda^3,\lambda^8$ are Gell-Mann matrices,
$\lambda^0$ is just in this context the $3\times3$ unit matrix.
These axial charges are related to the first moment of the polarized
quark distributions
\begin{equation}\label{Nucleon axial charges}
g^{(3)}_A=\Delta u-\Delta d,\qquad
g^{(8)}_A=\frac{1}{\sqrt{3}}(\Delta u+\Delta d-2\Delta s),\qquad
g^{(0)}_A=\Delta u+\Delta d+\Delta s
\end{equation}
where $\Delta q\equiv\int_0^1\ud
z\left[q_\uparrow(z)-q_\downarrow(z)+\bar q_\uparrow(z)-\bar
q_\downarrow(z)\right]$. Because of isospin symmetry, we expect that
$g^{(3)}_A$ is the same as the axial charge obtained by the matrix
element of the transition $p\to\pi^+ n$.

\subsection{3-quark contribution}

If one looks to the 3-quark component of a baryon wavefunction, one
can see that there are $3!$ possible and equivalent contractions of
the annihilation-creation operators. The contraction in color then
gives another factor of
$3!=\epsilon^{\alpha_1\alpha_2\alpha_3}\epsilon_{\alpha_1\alpha_2\alpha_3}$.
From eq. (\ref{T tensor},\ref{Normalization}) on can express the
normalization of the 3-quark component of baryon wavefunctions as
\begin{eqnarray}
\uN^{(3)}(B)&=&\frac{6\cdot
6}{2}\,\delta^k_lT(B)^{f_1f_2f_3}_{j_1j_2j_3,k}T(B)_{f_1f_2f_3}^{l_1l_2l_3,l}\int\ud
z_{1,2,3}\,\frac{\ud^2\up_{1,2,3\perp}}{(2\pi)^6}\,\delta(z_1+z_2+z_3-1)(2\pi)^2\delta^{(2)}(\up_{1\perp}+\up_{2\perp}+\up_{3\perp})\nonumber\\
&\times&F^{j_1\sigma_1}(p_1)F^{j_2\sigma_2}(p_2)F^{j_3\sigma_3}(p_3)F^\dag_{l_1\sigma_1}(p_1)F^\dag_{l_2\sigma_2}(p_2)F^\dag_{l_3\sigma_3}(p_3)\label{Normalization
3q}
\end{eqnarray}
where $F^{j\sigma}(p)\equiv F^{j\sigma}(z,\up_\perp)$ are the
discrete-level wavefunctions (\ref{Discrete level IMF},\ref{Discrete
level IMF 2}). In the non-relativistic limit, one can write
$F^{j\sigma}(p)F^\dag_{l\sigma}(p)\approx\delta^j_l h^2(p)$ (see eq.
(\ref{Approximation})). This 3-quark normalization is schematically
represented in Fig. \ref{Three-q normalization}.
\begin{figure}[h]\begin{center}\begin{minipage}[c]{8cm}\begin{center}\includegraphics[width=4cm]{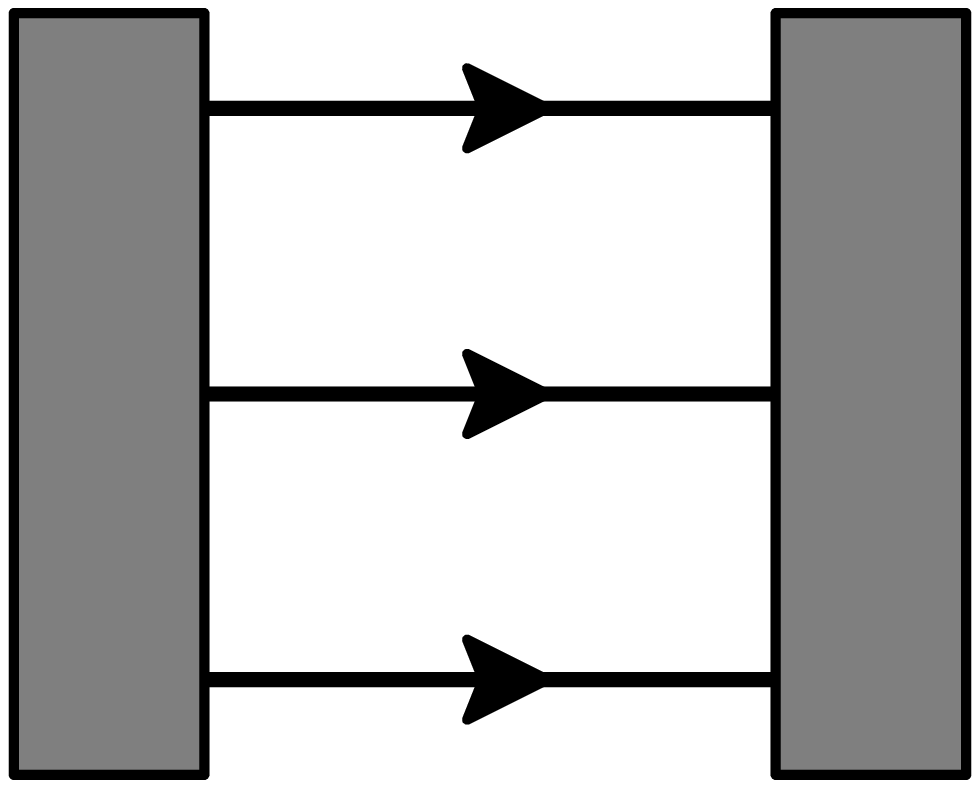}
\caption{\small{Schematic representation of the 3-quark
normalization. All contractions of the annihilation-creation
operators are equivalent to this specific one. Each quark line
stands for the color, flavor and spin contractions
$\delta^{\alpha_i}_{\alpha'_i}\delta^{f_i}_{f'_i}\delta^{\sigma_i}_{\sigma'_i}
\int\ud
z'_i\,\ud^2\up'_{i\perp}\delta(z_i-z'_i)\delta^{(2)}(\up_{i\perp}-\up'_{i\perp})$.}}\label{Three-q
normalization}
\end{center}\end{minipage}\hspace{0.5cm}
\begin{minipage}[c]{8cm}\begin{center}\includegraphics[width=4cm]{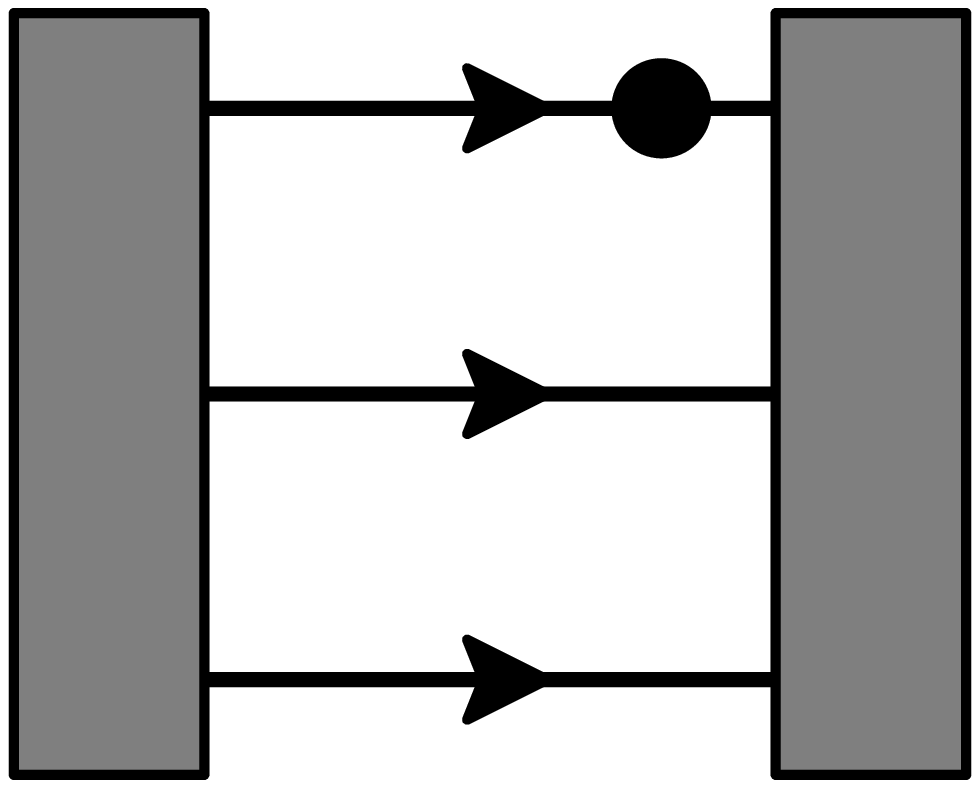}
\caption{\small{Schematic representation of the 3-quark contribution
to a charge. The black dot stands for the one-quark operator with
flavor content $J^e_h$. Since all three quark lines are equivalent
one has three times this specific contribution.}}\label{Three-q
charge}
\end{center}\end{minipage}\end{center}
\end{figure}

In the 3-quark sector, there is no antiquark which means that the
$b^\dag b$ part of the current does not play. As in the 3-quark
normalization one gets the factor $6\cdot 6$ from all contractions.
Let the third quark be the one whose charge is measured. One then
obtains an additional factor of 3 from the three quarks to which the
charge operator can be applied (see Fig. \ref{Three-q charge}). If
we denote by $\int(\ud p_{1-3})$ the integrals over momenta with the
conservation $\delta$-functions as in eq. (\ref{Normalization 3q})
one obtains the following expression for matrix element of the
vector charge
\begin{eqnarray}
V^{(3)}(1\to 2)&=&\frac{6\cdot 6\cdot
3}{2}\,\delta^k_lT(1)^{f_1f_2f_3}_{j_1j_2j_3,k}T(2)_{f_1f_2g_3}^{l_1l_2l_3,l}\int(\ud p_{1-3})\nonumber\\
&\times&\Big[F^{j_1\sigma_1}(p_1)F^{j_2\sigma_2}(p_2)F^{j_3\sigma_3}(p_3)\Big]\left[F^\dag_{l_1\sigma_1}(p_1)F^\dag_{l_2\sigma_2}(p_2)F^\dag_{l_3\tau_3}(p_3)\right]\left[\delta^{\tau_3}_{\sigma_3}J^{g_3}_{f_3}\right].\label{Vector
charge 3q}
\end{eqnarray}
We consider here for simplicity only matrix elements with zero
momentum transfer.

The axial charge is easily obtained from the vector one. One just
has to replace the averaging over baryon spin by
$\frac{1}{2}(-\sigma_3)^k_l$ and the axial charge operator involves
now $(-\sigma_3)^{\tau_3}_{\sigma_3}$ instead of
$\delta^{\tau_3}_{\sigma_3}$. One then has
\begin{eqnarray}
A^{(3)}(1\to 2)&=&\frac{6\cdot 6\cdot
3}{2}\,(-\sigma_3)^k_lT(1)^{f_1f_2f_3}_{j_1j_2j_3,k}T(2)_{f_1f_2g_3}^{l_1l_2l_3,l}\int(\ud p_{1-3})\nonumber\\
&\times&\Big[F^{j_1\sigma_1}(p_1)F^{j_2\sigma_2}(p_2)F^{j_3\sigma_3}(p_3)\Big]\left[F^\dag_{l_1\sigma_1}(p_1)F^\dag_{l_2\sigma_2}(p_2)F^\dag_{l_3\tau_3}(p_3)\right]\left[(-\sigma_3)^{\tau_3}_{\sigma_3}J^{g_3}_{f_3}\right].\label{Axial
charge 3q}
\end{eqnarray}

\subsection{5-quark contributions}\label{fivequarks diagrams}

\begin{figure}[h]\begin{center}\begin{minipage}[c]{10cm}\begin{center}\includegraphics[width=3cm]{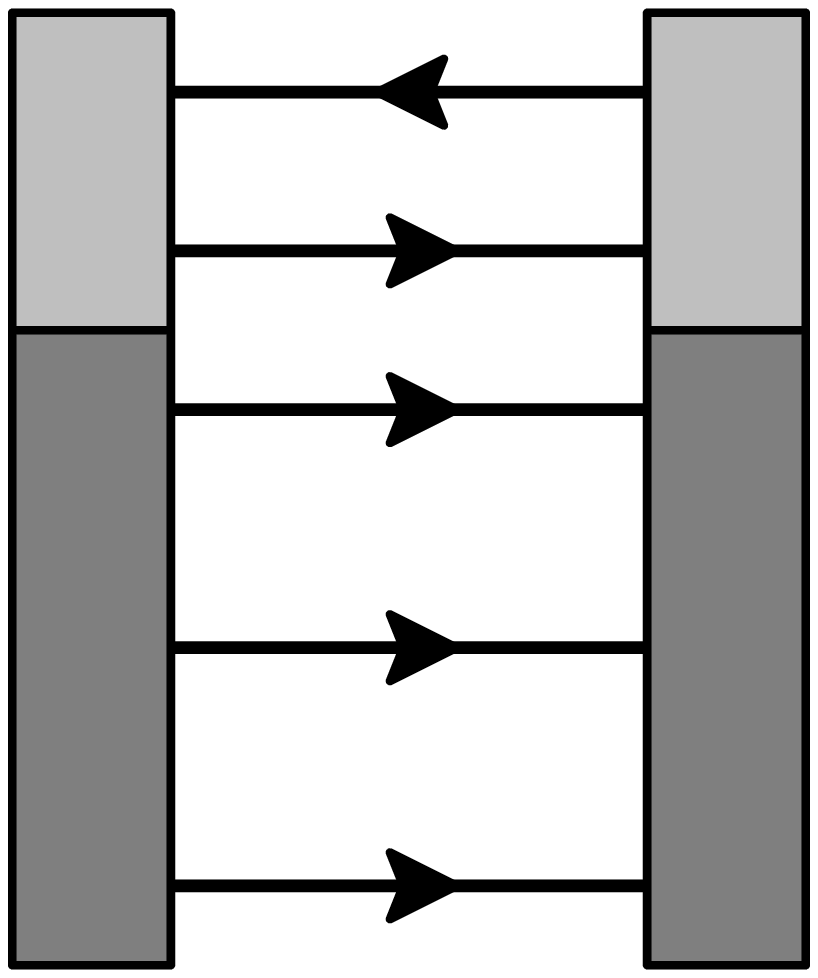}\hspace{1cm}\includegraphics[width=3cm]{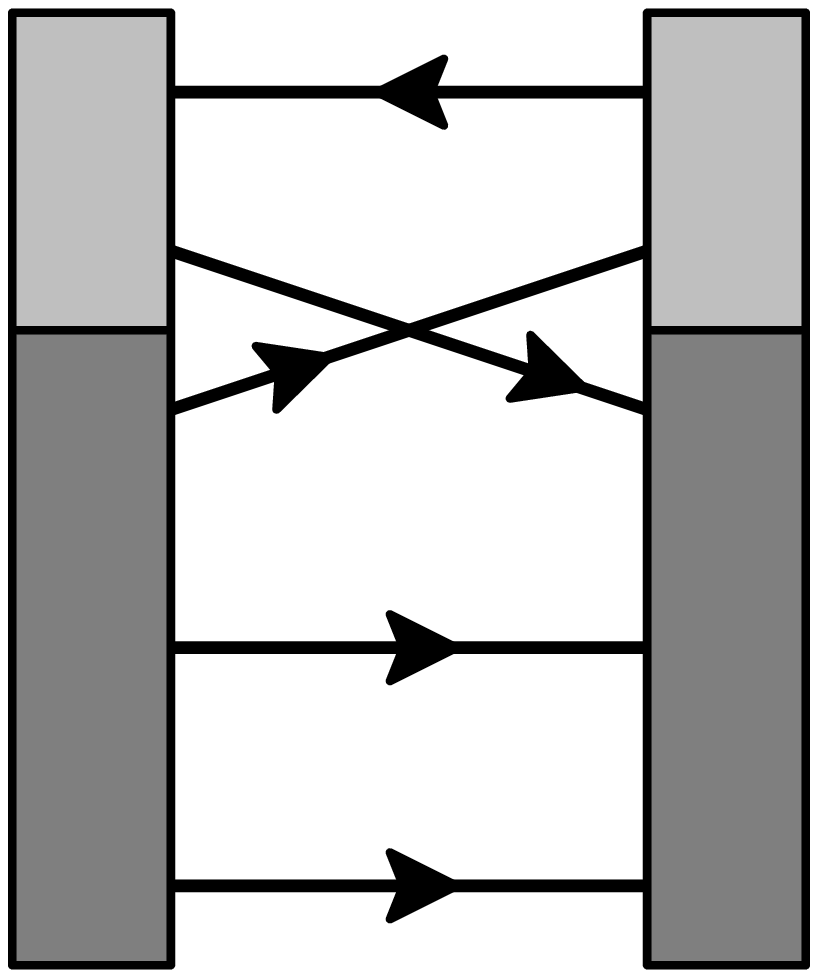}
\caption{\small{Schematic representation of the 5-quark direct
(left) and exchange (right) contributions to the
normalization.}}\label{Five-q normalization}
\end{center}\end{minipage}\end{center}\end{figure}

In the 5-quark component of the baryon wavefunctions there are
already two types of contributions to the normalization: the direct
and the exchange ones (see Fig. \ref{Five-q normalization}).
In the former, one contracts the $a^\dag$ from the pair wavefunction
with the $a$ in the conjugate pair and all the valence operators are
contracted with each other. As in the 3-quark normalization, there
are 6 equivalent possibilities but the contractions in color give
now a factor of $6\cdot
3=\epsilon^{\alpha_1\alpha_2\alpha_3}\epsilon_{\alpha_1\alpha_2\alpha_3}\delta^{\alpha}_{\alpha}$
because of the sum over color in the pair, then giving a total
factor of 108. In the exchange contribution, one contracts the
$a^\dag$ from the pair with one of the three $a$'s from the
conjugate discrete level. \emph{Vice versa}, the $a$ from the
conjugate pair is contracted with one of the three $a^\dag$'s from
the discrete level. There are at all 18 equivalent possibilities but
the contractions in color give only a factor of
$6=\epsilon^{\alpha_1\alpha_2\alpha}\epsilon_{\alpha_1\alpha_2\alpha_3}\delta^{\alpha_3}_{\alpha}$
and so one gets also a global factor of 108 for the exchange
contribution but with an additional minus sign because one has to
anticommute fermion operators to obtain exchange terms. We thus
obtain the following expression for the 5-quark normalization
\begin{eqnarray}
\uN^{(5)}(B)&=&\frac{108}{2}\,\delta^k_lT(B)^{f_1f_2f_3f_4,j_5}_{j_1j_2j_3j_4,f_5,k}T(B)_{f_1f_2g_3g_4,l_5}^{l_1l_2l_3l_4,f_5,l}\int(\ud p_{1-5})\nonumber\\
&\times&F^{j_1\sigma_1}(p_1)F^{j_2\sigma_2}(p_2)F^{j_3\sigma_3}(p_3)W^{j_4\sigma_4}_{j_5\sigma_5}(p_4,p_5)F^\dag_{l_1\sigma_1}(p_1)F^\dag_{l_2\sigma_2}(p_2)\nonumber\\
&\times&\left[F^\dag_{l_3\sigma_3}(p_3)W_{c\,l_4\sigma_4}^{l_5\sigma_5}(p_4,p_5)\delta^{g_3}_{f_3}\delta^{g_4}_{f_4}-F^\dag_{l_3\sigma_4}(p_4)W_{c\,l_4\sigma_3}^{l_5\sigma_5}(p_3,p_5)\delta^{g_3}_{f_4}\delta^{g_4}_{f_3}\right]\label{Normalization
5q}
\end{eqnarray}
where we have denoted
\begin{equation}
\int(\ud p_{1-5})=\int\ud
z_{1-5}\,\delta(z_1+\ldots+z_5-1)\int\frac{\ud^2\up_{1-5\perp}}{(2\pi)^{10}}\,(2\pi)^2\delta^{(2)}(\up_{1\perp}+\ldots+\up_{5\perp}).
\end{equation}
These schematic representations or diagrams are really useful when
one wishes to determine all the different possible contractions of
annihilation-creation operators, the number of equivalent ones and
their relative signs. In Appendix B we give some general rules that
help one that desires to explore any specific Fock component of a
baryon.

Concerning the vector and axial charges, we have three types of
direct contributions and four types of exchange contributions.
\begin{figure}[h]\begin{center}\begin{minipage}[c]{6.86cm}\begin{center}\includegraphics[width=6.86cm]{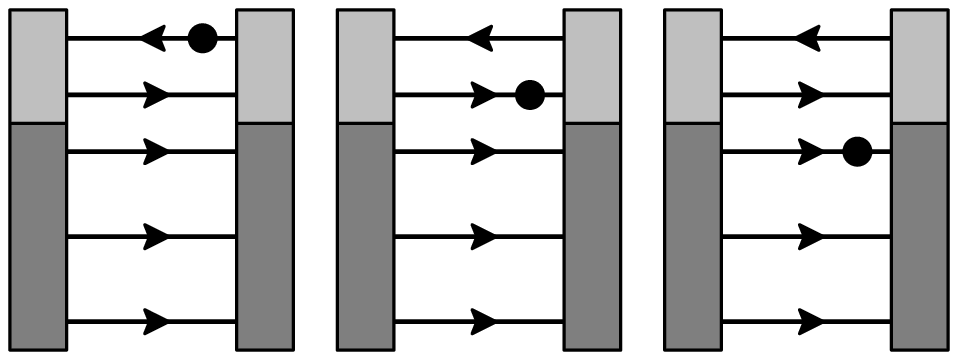}
\caption{\small{Schematic representation of the three type of
5-quark direct contributions to the charges.}}\label{Direct charges}
\end{center}\end{minipage}\hspace{0.5cm}
\begin{minipage}[c]{9.14cm}\begin{center}\includegraphics[width=9.14cm]{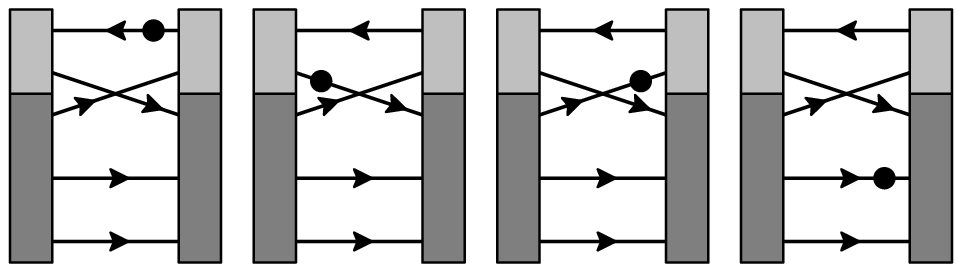}
\caption{\small{Schematic representation of the four types of
5-quark exchange contributions to the
charges.\newline}}\label{Exchange charges}
\end{center}\end{minipage}\end{center}
\end{figure}
From schematic representations of these contributions  (see Figs.
\ref{Direct charges},\ref{Exchange charges}), it is easy to write
the direct and exchange transitions. We will write only vector
charges since axial ones are obtained in the same way as in the
3-quark sector (the charge operator is in bold).\newline\newline
Direct contributions:
\begin{eqnarray}
V^{(5)\textrm{direct}}(1\to 2)&=&\frac{108}{2}\,\delta^k_lT(1)^{f_1f_2f_3f_4,j_5}_{j_1j_2j_3j_4,f_5,k}T(2)_{f_1f_2g_3g_4,l_5}^{l_1l_2l_3l_4,g_5,l}\int(\ud p_{1-5})\nonumber\\
&\times&F^{j_1\sigma_1}(p_1)F^{j_2\sigma_2}(p_2)F^{j_3\sigma_3}(p_3)W^{j_4\sigma_4}_{j_5\sigma_5}(p_4,p_5)F^\dag_{l_1\sigma_1}(p_1)F^\dag_{l_2\sigma_2}(p_2)F^\dag_{l_3\tau_3}(p_3)W_{c\,l_4\tau_4}^{l_5\tau_5}(p_4,p_5)\nonumber\\
&\times&\left[-\delta^{g_3}_{f_3}\delta^{g_4}_{f_4}\boldsymbol{
J^{f_5}_{g_5}}\delta^{\tau_3}_{\sigma_3}\delta^{\tau_4}_{\sigma_4}\boldsymbol{\delta_{\tau_5}^{\sigma_5}}+\delta^{g_3}_{f_3}\boldsymbol{
J^{g_4}_{f_4}}\delta^{f_5}_{g_5}\delta^{\tau_3}_{\sigma_3}\boldsymbol{\delta^{\tau_4}_{\sigma_4}}\delta_{\tau_5}^{\sigma_5}+3\boldsymbol{
J^{g_3}_{f_3}}\delta^{g_4}_{f_4}\delta^{f_5}_{g_5}\boldsymbol{\delta^{\tau_3}_{\sigma_3}}\delta^{\tau_4}_{\sigma_4}\delta_{\tau_5}^{\sigma_5}\right].\label{Direct}
\end{eqnarray}
Exchange contributions:
\begin{eqnarray}
V^{(5)\textrm{exchange}}(1\to 2)&=&-\frac{108}{2}\,\delta^k_lT(1)^{f_1f_2f_3f_4,j_5}_{j_1j_2j_3j_4,f_5,k}T(2)_{f_1g_2g_3g_4,l_5}^{l_1l_2l_3l_4,g_5,l}\int(\ud p_{1-5})\nonumber\\
&\times&F^{j_1\sigma_1}(p_1)F^{j_2\sigma_2}(p_2)F^{j_3\sigma_3}(p_3)W^{j_4\sigma_4}_{j_5\sigma_5}(p_4,p_5)F^\dag_{l_1\sigma_1}(p_1)F^\dag_{l_2\tau_2}(p_2)F^\dag_{l_3\tau_3}(p_3)W_{c\,l_4\tau_4}^{l_5\tau_5}(p_3,p_5)\nonumber\\
&\times&\left[-\delta^{g_2}_{f_2}\delta^{g_4}_{f_3}\delta^{g_3}_{f_4}\boldsymbol{
J^{f_5}_{g_5}}\delta^{\tau_2}_{\sigma_2}\delta^{\tau_4}_{\sigma_3}\delta^{\tau_3}_{\sigma_4}\boldsymbol{\delta_{\tau_5}^{\sigma_5}}+\delta^{g_2}_{f_2}\delta^{g_4}_{f_3}\boldsymbol{
J^{g_3}_{f_4}}\delta^{f_5}_{g_5}\delta^{\tau_2}_{\sigma_2}\delta^{\tau_4}_{\sigma_3}\boldsymbol{\delta^{\tau_3}_{\sigma_4}}\delta_{\tau_5}^{\sigma_5}\right.\nonumber\\
&&+\left.\delta^{g_2}_{f_2}\boldsymbol{
J^{g_4}_{f_3}}\delta^{g_3}_{f_4}\delta^{f_5}_{g_5}\delta^{\tau_2}_{\sigma_2}\boldsymbol{\delta^{\tau_4}_{\sigma_3}}\delta^{\tau_3}_{\sigma_4}\delta_{\tau_5}^{\sigma_5}
+2\boldsymbol{
J^{g_2}_{f_2}}\delta^{g_4}_{f_3}\delta^{g_3}_{f_4}\delta^{f_5}_{g_5}\boldsymbol{\delta^{\tau_2}_{\sigma_2}}\delta^{\tau_4}_{\sigma_3}\delta^{\tau_3}_{\sigma_4}\delta_{\tau_5}^{\sigma_5}\right].\label{Exchange}
\end{eqnarray}

We apply in the next sections these general formulae to compute the
nucleon axial charges and estimate the $\Theta^+$ width.

\section{Scalar overlap integrals in the IMF}\label{Section dix}

The contractions in eqs. (\ref{Normalization
5q},\ref{Direct},\ref{Exchange}) are easily performed by
\emph{Mathematica} over all flavor $(f,g)$, isospin $(j,l)$ and spin
$(\sigma,\tau)$ indices. One is then left with scalar integrals over
longitudinal $z$ and transverse $\up_\perp$ momenta of the five
quarks. The integrals over relative transverse momenta in the $q\bar
q$ pair are generally UV divergent. This divergence should be cut by
the momentum-dependent dynamical quark mass $M(p)$ (see eq.
(\ref{Effective lagrangian})). Following the authors of
\cite{Cutoff} we shall mimic the fall-off of $M(p)$ by the
Pauli-Villars cutoff at $M_\textrm{PV}=556.8$ MeV (this value being
chosen from the requirement that the pion decay constant $F_\pi=93$
MeV is reproduced from $M(0)=345$ MeV).

The pair wavefunction (\ref{Pair wavefunction}) is given in terms of
the Fourier transforms of the mean chiral field $\Pi(\uq)$ and
$\Sigma(\uq)$ (\ref{Fourier tranform of mean field}). One has
\begin{eqnarray}
&&\Pi(\uq)^j_{j'}=i\frac{(q^a\tau^a)^j_{j'}}{|\uq|}\,\Pi(q),\qquad\Pi(q)=\frac{4\pi}{q^2}\int_0^\infty\ud
r\,\sin P(r)(qr\cos qr-\sin qr)<0,\\
&&\Sigma(\uq)^j_{j'}=\delta^j_{j'}\Sigma(q),\qquad\Sigma(q)=\frac{4\pi}{q}\int_0^\infty\ud
r\,r(\cos P(r)-1)\sin qr<0.
\end{eqnarray}
We remind that $\uq$ is the 3-momentum of the $q\bar q$ pair which
is $\uq=((\up+\up')_\perp,(z+z')\uM)$.

\subsection{5-quark direct integrals (old result)}

Diakonov and Petrov have derived and computed the 5-quark direct
integrals. There are four of them where the quark-loop integrands
have to be understood as renormalized by the Pauli-Villars
prescription $G(y, \uQcal,\uq,M)- (M\to M_\textrm{PV})$
\begin{eqnarray}
K_{\pi\pi}&=&\frac{M^2}{2\pi}\int\frac{\ud^3\uq}{(2\pi)^3}\,\Phi\left(\frac{q_z}{\uM},\uq_\perp\right)\theta(q_z)q_z\Pi^2(\uq)\int^1_0\ud
y\int\frac{\ud^2\uQcal_\perp}{(2\pi)^2}\frac{\uQcal^2_\perp+M^2}{(\uQcal^2_\perp+M^2+y(1-y)\uq^2)^2},\label{Direct
begin}
\\
K_{\sigma\sigma}&=&\frac{M^2}{2\pi}\int\frac{\ud^3\uq}{(2\pi)^3}\,\Phi\left(\frac{q_z}{\uM},\uq_\perp\right)\theta(q_z)q_z\Sigma^2(\uq)\int^1_0\ud
y\int\frac{\ud^2\uQcal_\perp}{(2\pi)^2}\frac{\uQcal^2_\perp+M^2(2y-1)^2}{(\uQcal^2_\perp+M^2+y(1-y)\uq^2)^2},
\\
K_{33}&=&\frac{M^2}{2\pi}\int\frac{\ud^3\uq}{(2\pi)^3}\,\Phi\left(\frac{q_z}{\uM},\uq_\perp\right)\theta(q_z)\frac{q_z^3}{\uq^2}\Pi^2(\uq)\int^1_0\ud
y\int\frac{\ud^2\uQcal_\perp}{(2\pi)^2}\frac{\uQcal^2_\perp+M^2}{(\uQcal^2_\perp+M^2+y(1-y)\uq^2)^2},
\\
K_{3\sigma}&=&\frac{M^2}{2\pi}\int\frac{\ud^3\uq}{(2\pi)^3}\,\Phi\left(\frac{q_z}{\uM},\uq_\perp\right)\theta(q_z)\frac{q_z^2}{|\uq|}\Pi(\uq)\Sigma(\uq)\int^1_0\ud
y\int\frac{\ud^2\uQcal_\perp}{(2\pi)^2}\frac{\uQcal^2_\perp+M^2(2y-1)}{(\uQcal^2_\perp+M^2+y(1-y)\uq^2)^2}.\label{Direct
end}
\end{eqnarray}
The authors have used the following variables
\begin{equation}
y=\frac{z'}{z+z'},\qquad\uQcal_\perp=\frac{z\up'_\perp-z'\up_\perp}{z+z'}.
\end{equation}
This set of variables allows one to first integrate over the
relative momenta inside the $q\bar q$ pair $y$, $\uQcal_\perp$ and
then over the 3-momentum $\uq$ of the pair as a whole. The step
function $\theta(q_z)$ ensures that the longitudinal momentum
carried by the pair is positive in the IMF. $\Phi(z,\uq_\perp)$
stands for the probability that three valence quarks ``leave'' the
longitudinal fraction $z=z_4+z_5=q_z/\uM$ and the transverse
momentum $\uq_\perp=\up_{4\perp}+\up_{5\perp}$ to the $q\bar q$
pair. In the non-relativistic limit, one has
\begin{equation}\label{Probability3q}
\Phi(z,\uq_\perp)=\int\ud
z_{1,2,3}\frac{\ud^2\up_{1,2,3\perp}}{(2\pi)^6}\,\delta(z+z_1+z_2+z_3-1)(2\pi)^2\delta^{(2)}(\uq_\perp+\up_{1\perp}+\up_{2\perp}+\up_{3\perp})h^2(p_1)h^2(p_2)h^2(p_3).
\end{equation}

Since in the 3-quark component of baryons there is no additional
$q\bar q$ pair, all non-relativistic quantities in this sector are
proportional to $\Phi(0,0)$. The normalization of the discrete-level
wavefunction $h(p)$ being arbitrary, we choose it such that
$\Phi(0,0)=1$.

\subsection{Relativistic corrections to the discrete-level wavefunction (new result)}

As quoted in \cite{Original paper} the uncertainty associated to the
non-relativistic approximation is expected to be large. Indeed, they
have systematically used the first-order perturbation theory in
$1-\epsilon$ where $\epsilon=E_\textrm{lev}/M\sim 0.58$. They have
thus
\begin{itemize}
\item ignored the lower component of the valence wavefunction $j(r)$
\item ignored the distortion of the valence wavefunction by the sea
(see eq. (\ref{Discrete level IMF 2}))
\item used the approximate expression for the pair wavefunction (see eq. (\ref{Pair wavefunction}))
\item neglected the 5-quark exchange diagrams when evaluating the
5-quark normalization and transition matrix elements
\item neglected the 7-, 9-, \ldots quark components in baryons.
\end{itemize}
There are three hints that this non-relativistic approximation is
not satisfactory: first the actual expansion parameter
$1-\epsilon=0.42$ is poor and second the ratio of the 5- to 3-quark
normalization is 50\%. Finally this can also be seen from the actual
components $h(r)$ and $j(r)$ of the discrete-level wavefunction
(Fig. \ref{Level}). Diakonov and Petrov commented that the lower
component $j(r)$ is ``substantially'' smaller than the upper one
$h(r)$. In fact the $j(r)$ contribution to the normalization of the
discrete-level wavefunction $\psi_\textrm{lev}(\ux)$ is still 20\%
(result in accordance with \cite{Lower component}). This combined
with combinatorics factors in eq. (\ref{Phi}) shows that considering
the lower component $j(r)$ can have a big impact on the estimations.
The nucleon is thus definitely a relativistic system.

We have improved the technique by considering the full expression
for the discrete-level wavefunction (\ref{Discrete level IMF}). We
have found that we have to use in the probability distribution
(\ref{Probability3q}) instead of $h^2(p_1)h^2(p_2)h^2(p_3)$ the
following combination
\begin{eqnarray}
&h^2(p_1)h^2(p_2)h^2(p_3)+6h^2(p_1)h^2(p_2)\left[h(p_3)\frac{p_{3z}}{|\up_3|}j(p_3)\right]+3h^2(p_1)h^2(p_2)j^2(p_3)&\nonumber\\
&+12h^2(p_1)\left[h(p_2)\frac{p_{2z}}{|\up_2|}j(p_2)\right]\left[h(p_3)\frac{p_{3z}}{|\up_3|}j(p_3)\right]+12h^2(p_1)\left[h(p_2)\frac{p_{2z}}{|\up_2|}j(p_2)\right]j^2(p_3)&\nonumber\\
&+8\left[h(p_1)\frac{p_{1z}}{|\up_1|}j(p_1)\right]\left[h(p_2)\frac{p_{2z}}{|\up_2|}j(p_2)\right]\left[h(p_3)\frac{p_{3z}}{|\up_3|}j(p_3)\right]+3h^2(p_1)j^2(p_2)j^2(p_3)&\nonumber\\
&+12\left[h(p_1)\frac{p_{1z}}{|\up_1|}j(p_1)\right]\left[h(p_2)\frac{p_{2z}}{|\up_2|}j(p_2)\right]j^2(p_3)+6\left[h(p_1)\frac{p_{1z}}{|\up_1|}j(p_1)\right]j^2(p_2)j^2(p_3)+j^2(p_1)j^2(p_2)j^2(p_3)\label{Phi}
\end{eqnarray}
where of course $p_{iz}=z_i\uM-E_\textrm{lev}$. When an axial
operator acts on the valence quarks it sees a slightly different
probability distribution (this integral will be denoted by
$\Psi(z,\uq_\perp)$)
\begin{eqnarray}
&h^2(p_1)h^2(p_2)h^2(p_3)+6h^2(p_1)h^2(p_2)\left[h(p_3)\frac{p_{3z}}{|\up_3|}j(p_3)\right]+h^2(p_1)h^2(p_2)\frac{2p_{3z}^2+p_3^2}{p_3^2}j^2(p_3)&\nonumber\\
&+12h^2(p_1)\left[h(p_2)\frac{p_{2z}}{|\up_2|}j(p_2)\right]\left[h(p_3)\frac{p_{3z}}{|\up_3|}j(p_3)\right]+4h^2(p_1)\left[h(p_2)\frac{p_{2z}}{|\up_2|}j(p_2)\right]\frac{2p_{3z}^2+p_3^2}{p_3^2}j^2(p_3)&\nonumber\\
&+8\left[h(p_1)\frac{p_{1z}}{|\up_1|}j(p_1)\right]\left[h(p_2)\frac{p_{2z}}{|\up_2|}j(p_2)\right]\left[h(p_3)\frac{p_{3z}}{|\up_3|}j(p_3)\right]+h^2(p_1)j^2(p_2)\frac{4p_{3z}^2-p_3^2}{p_3^2}j^2(p_3)&\nonumber\\
&+4\left[h(p_1)\frac{p_{1z}}{|\up_1|}j(p_1)\right]\left[h(p_2)\frac{p_{2z}}{|\up_2|}j(p_2)\right]\frac{2p_{3z}^2+p_3^2}{p_3^2}j^2(p_3)+2\left[h(p_1)\frac{p_{1z}}{|\up_1|}j(p_1)\right]j^2(p_2)\frac{4p_{3z}^2-p_3^2}{p_3^2}j^2(p_3)&\nonumber\\
&+j^2(p_1)j^2(p_2)\frac{2p_{3z}^2-p_3^2}{p_3^2}j^2(p_3).&\label{Psi}
\end{eqnarray}
This distribution has been normalized in such a way that the
prefactor of the axial charge is the same as the one of the vector
charge (\ref{Direct}).

Then in the 3-quark component of baryons all quantities are
proportional to either $\Phi(0,0)$ or $\Psi(0,0)$. The normalization
of the discrete-level wavefunctions $h(p)$ and $j(p)$ being
arbitrary, we choose it such that $\Phi(0,0)=1$.

Note that we still haven't taken into account the distortion of the
valence level due to the sea.

\subsection{5-quark exchange integrals (new result)}

Our other improvement of the technique is the consideration of the
exchange diagrams which were believed to have a strong impact on
observables because of their sign opposite to the direct one
\cite{Original paper} (see for example eq. (\ref{Normalization
5q})). We have found that for the exchange contributions there were
thirteen non-zero scalar integrals. Since the quark from the sea is
exchanged with a valence quark, we cannot disentangle the
quark-antiquark pair from the valence quarks. At best two valence
quarks can be factorized out and leave 9-dimensional integrals
\begin{equation}
K=\frac{M^2}{2\pi}\int(\ud p_{3,4,5})
\,\phi\left(\uZ,\uP_\perp\right)\frac{\uM^2}{2\pi
Z'Z}\,I(z_{3,4,5},\up_{3,4,5\perp})h(p_3)h(p_4),
\end{equation}
where $\uZ=z_3+z_4+z_5$, $\uP_\perp=(\up_3+\up_4+\up_5)_\perp$, $Z$
is given by eq. (\ref{Notation}) with $z=z_4$ and $z'=z_5$ while
$Z'$ is the same but with the replacement $z_4\to z_3$. The function
$I(z_{3,4,5},\up_{3,4,5\perp})$ stands for the thirteen integrands
\begin{eqnarray}
I_1&=&\Sigma(\uq')\Sigma(\uq)\left(\uQ'_\perp\cdot
\uQ_\perp+M^2(z_5-z_3)(z_5-z_4)\right),\label{Exchange begin}\\
I_2&=&\Pi(\uq')\Pi(\uq)\,\frac{q'\cdot q}{q'q}\left(\uQ'_\perp\cdot
\uQ_\perp+M^2(z_5+z_3)(z_5+z_4)\right),\\
I_3&=&\Pi(\uq')\Pi(\uq)\,\frac{\uq'_\perp\times
\uq_\perp}{q'q}\,(\uQ'_\perp\times
\uQ_\perp),\\
I_4&=&\Pi(\uq')\Pi(\uq)\,\frac{M(\uq'_\perp q_z-\uq_\perp
q'_z)}{q'q}\cdot\left(\uQ_\perp-
\uQ'_\perp\right),\\
I_5&=&\Pi(\uq')\Pi(\uq)\,\frac{q'_zq_z}{q'q}\left(\uQ'_\perp\cdot
\uQ_\perp+M^2(z_5+z_3)(z_5+z_4)\right),\\
I_6&=&\Sigma(\uq')\Pi(\uq)\,\frac{q_z}{q}\left(\uQ'_\perp\cdot
\uQ_\perp+M^2(z_5-z_3)(z_5+z_4)\right),\\
I_7&=&\Sigma(\uq')\Pi(\uq)\,\frac{M\uq_\perp}{q}\cdot\left(\uQ'_\perp
(z_5+z_4)-\uQ_\perp(z_5-z_3)\right),\\
I_8&=&\Sigma(\uq')\Sigma(\uq)\left(\uQ'_\perp\cdot
\uQ_\perp-M^2(z_5-z_3)(z_5-z_4)\right),\\
I_9&=&\Pi(\uq')\Pi(\uq)\,\frac{q'\cdot q}{q'q}\left(\uQ'_\perp\cdot
\uQ_\perp-M^2(z_5+z_3)(z_5+z_4)\right),\\
I_{10}&=&\Pi(\uq')\Pi(\uq)\,\frac{M(\uq'_\perp q_z+\uq_\perp
q'_z)}{q'q}\cdot\left(\uQ_\perp+ \uQ'_\perp\right),\\
I_{11}&=&\Pi(\uq')\Pi(\uq)\,\frac{q'_zq_z}{q'q}\left(\uQ'_\perp\cdot
\uQ_\perp-M^2(z_5+z_3)(z_5+z_4)\right),\\
I_{12}&=&\Sigma(\uq')\Pi(\uq)\,\frac{q_z}{q}\left(\uQ'_\perp\cdot
\uQ_\perp-M^2(z_5-z_3)(z_5+z_4)\right),\\
I_{13}&=&\Sigma(\uq')\Pi(\uq)\,\frac{M\uq_\perp}{q}\cdot\left(\uQ'_\perp
(z_5+z_4)+\uQ_\perp(z_5-z_3)\right),\label{Exchange end}
\end{eqnarray}
where $\uq=((\up_4+\up_5)_\perp,(z_4+z_5)\uM)$ and
$\uQ_\perp=z_4\up_{5\perp}-z_5\up_{4\perp}$. The primed variables
stand for the same as the unprimed ones but with the replacement
$z_4\to z_3$. The regularization of those integrals is done exactly
in the same way as for the direct contributions.

The function $\phi(\uZ,\uP_\perp)$ stands for the probability that
two valence quarks ``leave'' the longitudinal fraction
$\uZ=z_3+z_4+z_5$ and the transverse momentum
$\uP_\perp=\up_{3\perp}+\up_{4\perp}+\up_{5\perp}$ to the rest of
the partons
\begin{equation}\label{probability2q}
\phi(\uZ,\uP_\perp)=\int\ud
z_{1,2}\frac{\ud^2\up_{1,2\perp}}{(2\pi)^4}\,\delta(\uZ+z_1+z_2-1)(2\pi)^2\delta^{(2)}(\uP_\perp+\up_{1\perp}+\up_{2\perp})h^2(p_1)h^2(p_2).
\end{equation}
We have kept of course the same normalization of the discrete-level
wavefunction $h(p)$ as in the direct contributions, i.e. such that
$\Phi(0,0)=\int(\ud p)\,\phi(z,\up_\perp)h^2(p)=1$. Anticipating on
the results, we haven't considered relativistic corrections to this
probability distribution since exchange contributions appear to be
fairly negligible. Exchange contributions have then been computed
only in the non-relativistic limit.

\section{Results}\label{Section onze}

All normalizations, vector and axial charges are linear combinations
of (\ref{Direct begin})-(\ref{Direct end}) for the direct
contributions and of (\ref{Exchange begin})-(\ref{Exchange end}) for
the exchange ones.

\subsection{Old results}

In their paper \cite{Original paper}, Diakonov and Petrov have
obtained the following combinations
\newline\newline Nucleon
normalization:
\begin{eqnarray}
\uN^{(3)}(N)&=&9\,\Phi(0,0),\\
\uN^{(5)\textrm{direct}}(N)&=&\frac{18}{5}\left(11K_{\pi\pi}+23K_{\sigma\sigma}\right).
\end{eqnarray}
Axial charge of the $p\to\pi^+n$ transition:
\begin{eqnarray}
A^{(3)}(p\to\pi^+n)&=&15\,\Phi(0,0),\\
A^{(5)\textrm{direct}}(p\to\pi^+n)&=&\frac{6}{25}\left(209K_{\pi\pi}+559K_{\sigma\sigma}-34K_{33}-356K_{3\sigma}\right).
\end{eqnarray}
$\Theta^+$ normalization:
\begin{equation}
\uN^{(5)\textrm{direct}}(\Theta)=\frac{36}{5}\left(K_{\pi\pi}+K_{\sigma\sigma}\right).
\end{equation}
Axial charge of the $\Theta^+\to K^+n$ transition:
\begin{equation}
A^{(5)\textrm{direct}}(\Theta^+\to
K^+n)=\frac{6}{5}\sqrt{\frac{3}{5}}\left(-7K_{\pi\pi}-5K_{\sigma\sigma}+8K_{33}+28K_{3\sigma}\right).
\end{equation}

\subsection{New results}

We have obtained the exchange combinations relative to these
quantities. On the top of that we have computed the matrix elements
of $\bar q\gamma_0\gamma_5q$ with $q=u,d,s$ for the nucleon in order
to obtain the three nucleon axial charges (\ref{Nucleon axial
charges}).\newline\newline Nucleon normalization:
\begin{equation}
\uN^{(5)\textrm{exchange}}(N)=\frac{-12}{5}\left(9K_1+4K_3+4K_4-17K_6-17K_7\right).
\end{equation}
Axial charge of the $p\to\pi^+n$ transition:
\begin{eqnarray}
A^{(5)\textrm{exchange}}(p\to\pi^+n)&=&\frac{-2}{25}\left(557K_1+K_2+221K_3+192K_4-2K_5-908K_6-978K_7\right.\nonumber\\
&&\left.+98K_8-50K_9+62K_{10}+124K_{11}-48K_{12}-100K_{13}\right).
\end{eqnarray}
Proton first moment of polarized quark distributions:
\begin{eqnarray}
\Delta u^{(3)}&=&12\,\Phi(0,0),\\
\Delta d^{(3)}&=&-3\,\Phi(0,0),\\
\Delta s^{(3)}&=&0,\\
\Delta
u^{(5)\textrm{direct}}(p)&=&\frac{18}{25}\left(41K_{\pi\pi}+151K_{\sigma\sigma}+14K_{33}-74K_{3\sigma}\right),\\
\Delta
d^{(5)\textrm{direct}}(p)&=&\frac{12}{25}\left(-43K_{\pi\pi}-53K_{\sigma\sigma}+38K_{33}+67K_{3\sigma}\right),\\
\Delta
s^{(5)\textrm{direct}}(p)&=&\frac{12}{25}\left(-11K_{\pi\pi}-K_{\sigma\sigma}+16K_{33}+14K_{3\sigma}\right),\end{eqnarray}
\begin{eqnarray}
\Delta
u^{(5)\textrm{exchange}}(p)&=&\frac{-6}{25}\left(153K_1-K_2+49K_3+48K_4+2K_5-262K_6-232K_7\right.\nonumber\\
&&\left.+32K_8+8K_{10}+16K_{11}-32K_{12}\right),\\
\Delta
d^{(5)\textrm{exchange}}(p)&=&\frac{-4}{25}\left(-49K_1-2K_2-37K_3-24K_4+4K_5+61K_6+141K_7\right.\nonumber\\
&&\left.-K_8+25K_9-19K_{10}-38K_{11}-24K_{12}+50K_{13}\right),\\
\Delta
s^{(5)\textrm{exchange}}(p)&=&\frac{-2}{25}\left(14K_1-8K_2-13K_3-6K_4+16K_5-26K_6+54K_7\right.\nonumber\\
&&\left.+11K_8+25K_9-16K_{10}-32K_{11}-36K_{12}+50K_{13}\right).
\end{eqnarray}
It is then easy to obtain the three axial charges. As expected by
isospin symmetry the axial charge obtained by the $p\to\pi^+n$
transition is the same as $g_A^{(3)}$ in any of the 3- or 5-quark
direct or exchange contributions
\begin{eqnarray}
g^{(3)}_{A^{(3)}}&=&A^{(3)}(p\to\pi^+n),\\
g^{(8)}_{A^{(3)}}&=&3\sqrt{3}\,\Phi(0,0),\\
g^{(0)}_{A^{(3)}}&=&9\,\Phi(0,0),\\
g^{(3)}_{A^{(5)\textrm{direct}}}&=&A^{(5)\textrm{direct}}(p\to\pi^+n),\\
g^{(8)}_{A^{(5)\textrm{direct}}}&=&\frac{18\sqrt{3}}{25}\left(9K_{\pi\pi}+39K_{\sigma\sigma}+6K_{33}-16K_{3\sigma}\right),\\
g^{(0)}_{A^{(5)\textrm{direct}}}&=&\frac{18}{5}\left(K_{\pi\pi}+23K_{\sigma\sigma}+10K_{33}-4K_{3\sigma}\right),\\
g^{(3)}_{A^{(5)\textrm{exchange}}}&=&A^{(5)\textrm{exchange}}(p\to\pi^+n),\\
g^{(8)}_{A^{(5)\textrm{exchange}}}&=&\frac{-6\sqrt{3}}{25}\left(37K_1+K_2+11K_3+12K_4-2K_5-68K_6-58K_7\right.\nonumber\\
&&\left.+8K_8+2K_{10}+4K_{11}-8K_{12}\right),\\
g^{(0)}_{A^{(5)\textrm{exchange}}}&=&\frac{-6}{5}\left(25K_1-K_2+4K_3+6K_4+2K_5-46K_6-24K_7\right.\nonumber\\
&&\left.+7K_8+5K_9-2K_{10}-4K_{11}-12K_{12}+10K_{13}\right).
\end{eqnarray}
For the vector charge of the $p\to\pi^+n$ transition one gets
exactly the same expression as the normalization of the contribution
under consideration, which means that the vector charge is conserved
in each Fock component separately and even in the direct and
exchange sectors separately.

Here are our results for the $\Theta^+$ pentaquark\newline\newline
$\Theta^+$ normalization:
\begin{equation}
\uN^{(5)\textrm{exchange}}(\Theta)=\frac{-12}{5}\left(K_3+K_4-2K_6-2K_7\right).
\end{equation}
Axial charge of the $\Theta^+\to K^+n$ transition:
\begin{eqnarray}
A^{(5)\textrm{exchange}}(\Theta^+\to
K^+n)&=&\frac{-2}{5}\sqrt{\frac{3}{5}}\left(-7K_1+K_2-7K_3-3K_4-2K_5+4K_6+18K_7\right.\nonumber\\
&&\left.-10K_8+10K_9-10K_{10}-20K_{11}+20K_{13}\right).
\end{eqnarray}
When relativistic effects are considered, the axial operator changes
the structure of the probability distribution. One has then to
replace $K_{\pi\pi}$, $K_{\sigma\sigma}$ and $K_{33}$ by
$K_{\pi\pi}'$, $K_{\sigma\sigma}'$ and $K_{33}'$, i.e. the same
integrals but with $\Phi(z,\uq_\perp)$ (eq. (\ref{Phi})) replaced by
$\Psi(z,\uq_\perp)$ (eq. (\ref{Psi})). Note that $K_{3\sigma}$ is
not affected since this integral appears only when the axial
operator acts on the pair.

The numerical value of these matrix elements has to be properly
normalized as in the following example
\begin{equation}
g_A(\Theta\to KN)=\frac{A^{(5)\textrm{direct}}(\Theta^+\to
K^+n)+A^{(5)\textrm{exchange}}(\Theta^+\to
K^+n)}{\sqrt{\uN^{(5)\textrm{direct}}(\Theta)+\uN^{(5)\textrm{exchange}}(\Theta)}\sqrt{\uN^{(3)}(N)+\uN^{(5)\textrm{direct}}(N)+\uN^{(5)\textrm{exchange}}(N)}}.
\end{equation}

\section{Numerical results}\label{Section douze}

\begin{figure}[h]\begin{center}\begin{minipage}[c]{5.3cm}\begin{center}\includegraphics[width=5.3cm]{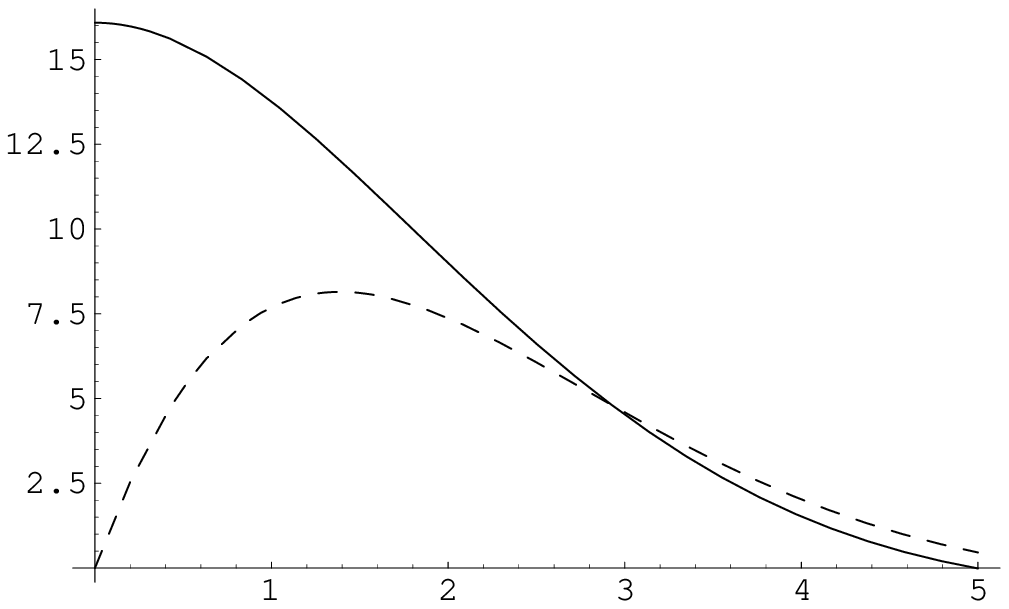}
\caption{\small{The self-consistent pseudoscalar $-|\uq|\Pi(\uq)$
(solid) and scalar $-|\uq|\Sigma(\uq)$ (dashed) fields in baryons.
The horizontal axis unit is $M$.}}\label{Self-consistent field plot}
\end{center}\end{minipage}\hspace{0.25cm}\begin{minipage}[c]{10.85cm}\begin{minipage}[c]{5.3cm}\begin{center}\includegraphics[width=5.3cm]{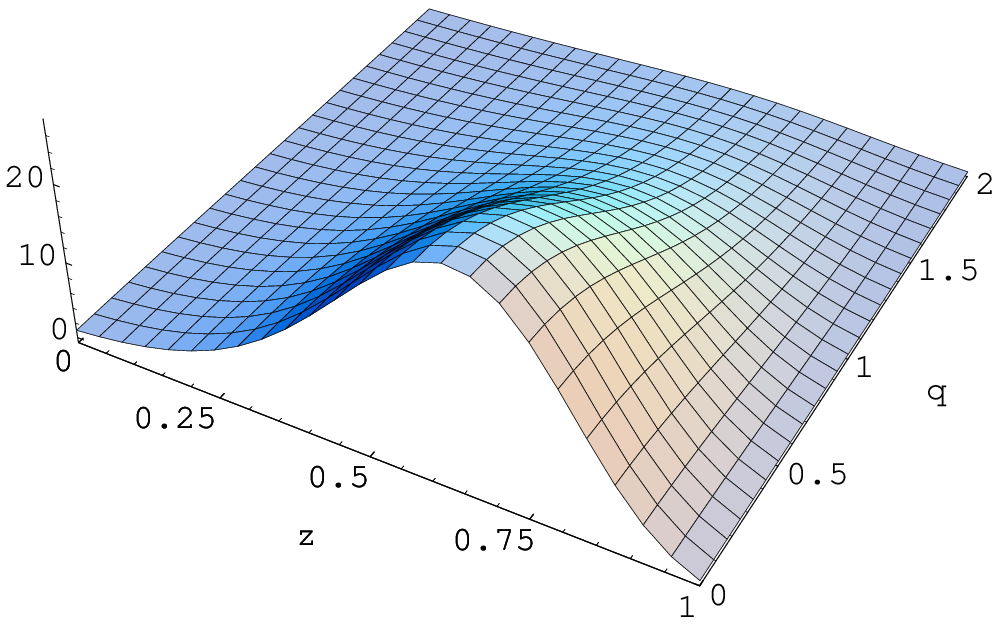}
\end{center}\end{minipage}\hspace{0.25cm}
\begin{minipage}[c]{5.3cm}\begin{center}\includegraphics[width=5.3cm]{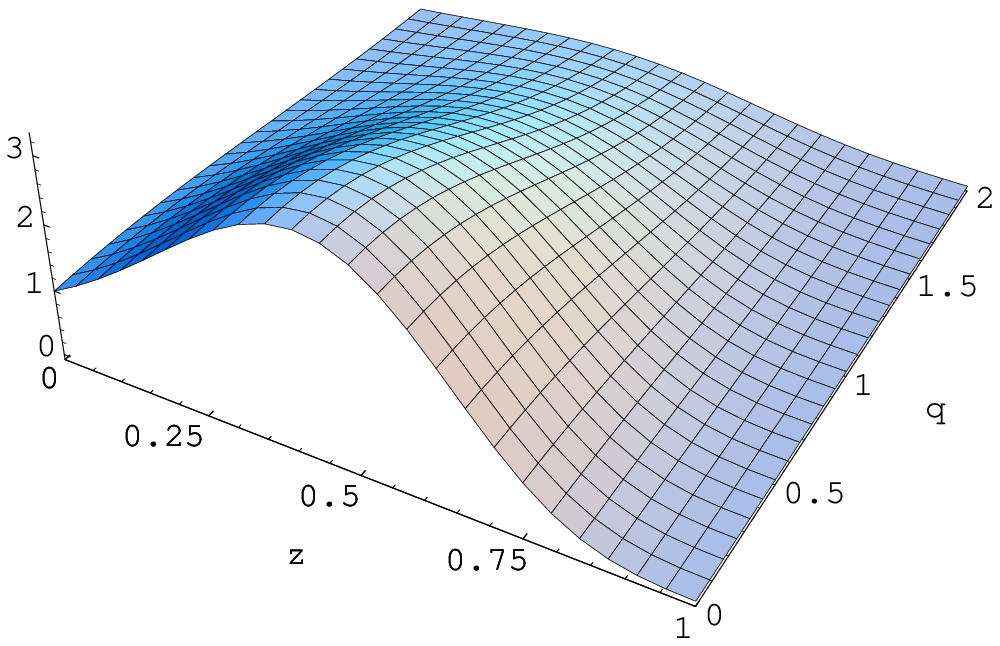}
\end{center}\end{minipage}\caption{\small{The non-relativistic probability distribution that two (left) or three (right) valence quarks leave the fraction $z$ of the baryon momentum and the
transverse momentum $\uq_\perp$ plotted in units of $M$ and
normalized to unity for
$z=\uq_\perp=0$.\newline\newline}}\label{Phiplot}\end{minipage}\end{center}
\end{figure}

\begin{figure}[h]\begin{center}\begin{minipage}[c]{5.3cm}\begin{center}\includegraphics[width=5.3cm]{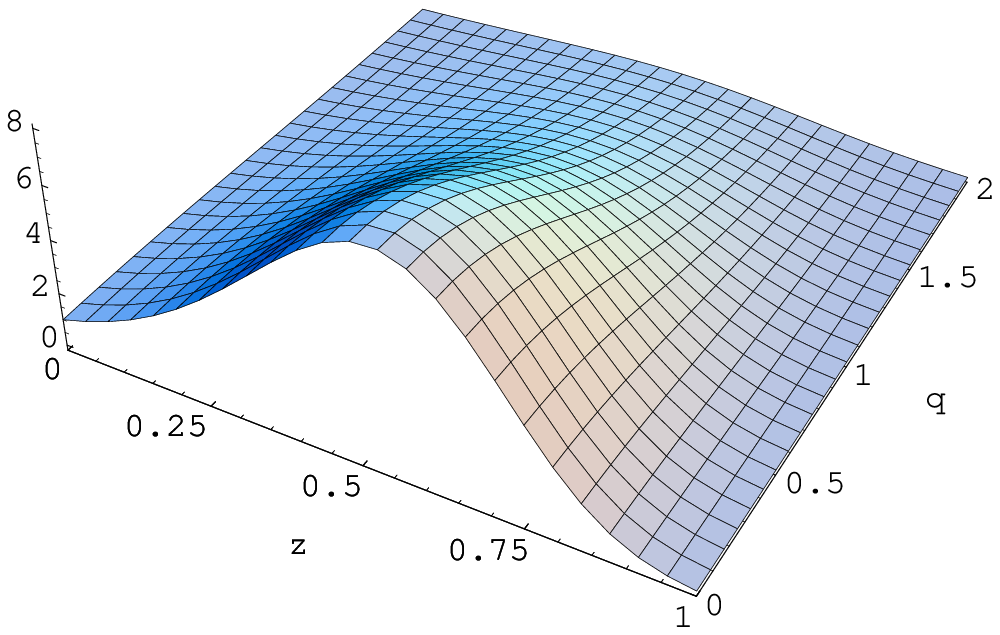}
\end{center}\end{minipage}\hspace{0.25cm}
\begin{minipage}[c]{5.3cm}\begin{center}\includegraphics[width=5.3cm]{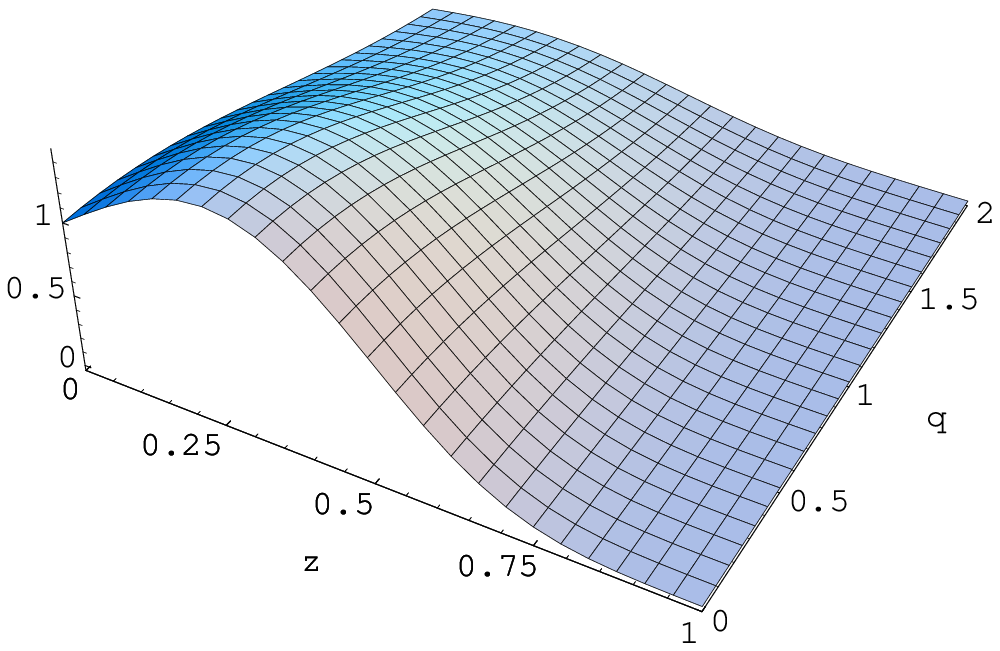}
\end{center}\end{minipage}\hspace{0.25cm}\begin{minipage}[c]{5.3cm}\begin{center}\includegraphics[width=5.3cm]{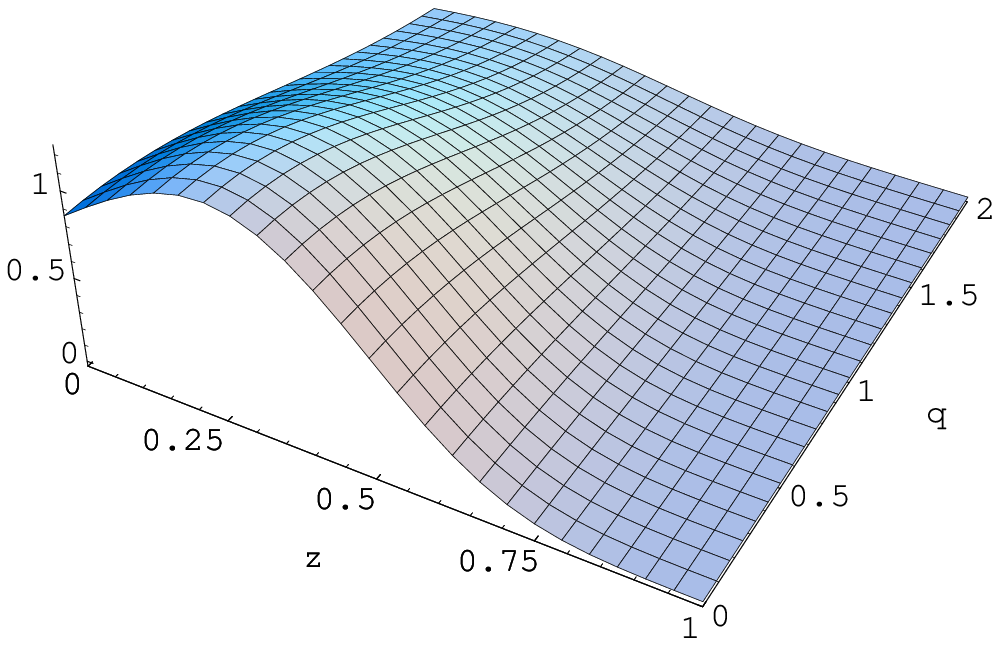}
\end{center}\end{minipage}\caption{\small{The probability distribution that two (left) or three (middle) valence quarks leave the fraction $z$ of the baryon momentum and the
transverse momentum $\uq_\perp$ with relativistic corrections to the
discrete-level wavefunction plotted in units of $M$ and normalized
to unity for $z=\uq_\perp=0$. Relativistic corrections clearly shift
the bump in the probability distributions to smaller values $z$
meaning that they leave less longitudinal momentum fraction to the
quark-antiquark pair. They seem also to smear a little bit this
bump. On the right is plotted the probability distribution that
enters scalar integrals when an axial charge is
considered.}}\label{Phiplot2}\end{center}
\end{figure}

In the evaluation of the scalar integrals we have used the quark
mass $M=345$ MeV, the self-consistent profile function (\ref{Profile
function}), the Pauli-Villars mass $M_\textrm{PV}=556.8$ MeV for the
regularization of (\ref{Direct begin})-(\ref{Direct end}),
(\ref{Exchange begin})-(\ref{Exchange end}) and the baryon mass
$\uM=1207$ MeV as it follows for the ``classical'' mass in the mean
field approximation \cite{Approximation}. The self-consistent scalar
$\Sigma(\uq)$ and pseudoscalar $\Pi(\uq)$ fields are plotted in Fig.
\ref{Self-consistent field plot}. The probability distributions
$\phi(z,\uq_\perp)$ (\ref{probability2q}) and $\Phi(z,\uq_\perp)$
(\ref{Probability3q}) that two or three valence quarks leave the
fraction $z$ of the baryon momentum and the transverse momentum
$\uq_\perp$ are plotted in Fig. \ref{Phiplot} in the
non-relativistic limit and in Fig. \ref{Phiplot2} with relativistic
corrections to the discrete-level wavefunction. By comparison one
immediately sees that relativistic corrections shift the bump in the
probability distributions to lower values of $z$ and smear it a
little bit. When relativistic corrections to an axial charge are
considered one has to use the $\Psi(z,\uq_\perp)$ probability
distribution which is slightly different (see Fig. \ref{Phiplot2})
from the relativistically corrected $\Phi(z,\uq_\perp)$. We remind
that the normalization of the discrete-level wavefunctions $h(p)$
(and $j(p)$) is chosen such that we have $\Phi(0,0)=1$.

The numerical evaluation of the non-relativistic direct integrals
(\ref{Direct begin})-(\ref{Direct end}) yields
\begin{equation}\label{Direct values}
K_{\pi\pi}=0.0624,\qquad K_{\sigma\sigma}=0.0284,\qquad
K_{33}=0.0373,\qquad K_{3\sigma}=0.0334.
\end{equation}
We have recalculated the integrals. The numerical precision is the
reason why these numbers are slightly different from those given in
\cite{Original paper}.

The numerical evaluation of the direct integrals (\ref{Direct
begin})-(\ref{Direct end}) with relativistic corrections to the
discrete-level wavefunction yields
\begin{equation}\label{Direct values rel}
K_{\pi\pi}=0.0365,\qquad K_{\sigma\sigma}=0.0140,\qquad
K_{33}=0.0197,\qquad K_{3\sigma}=0.0163.
\end{equation}
As one can expect from the comparison between Fig. \ref{Phiplot} and
\ref{Phiplot2} relativistic corrections reduce strongly (about one
half) the values of the scalar integrals.

The numerical evaluation of the direct integrals (\ref{Direct
begin})-(\ref{Direct end}) with relativistic corrections to the
discrete-level wavefunction that enter axial charges and first
moment of polarized quark distributions yields
\begin{equation}\label{Direct values axial rel}
K_{\pi\pi}'=0.0300,\qquad K_{\sigma\sigma}'=0.0112,\qquad
K_{33}'=0.0163.
\end{equation}

The numerical evaluation of the exchange integrals (\ref{Exchange
begin})-(\ref{Exchange end}) yields
\begin{eqnarray}
&K_1=0.0056,\qquad K_2=0.0097,\qquad K_3=-0.0008\qquad K_4=0.0047,&\nonumber\\
&K_5=0.0086,\qquad K_6=0.0042,\qquad K_7=0.0029,\qquad K_8=0.0043,\qquad K_9=0.0031,&\nonumber\\
&K_{10}=0.0069,\qquad K_{11}=0.0017,\qquad K_{12}=0.0057,\qquad
K_{13}=0.0023.&\label{Exchange values}
\end{eqnarray}

All nucleon axial charges and first moment of polarized quark
distributions are collected and presented in Table \ref{Nucleon
results}.
\begin{table}[h!]\begin{center}\caption{\small{Results for the nucleon: axial charges, first moment of polarized quark distributions and ratio of the 5- to the 3-quark normalization. First, results in the non-relativistic approximation
are given, then with relativistic corrections to the discrete-level
wavefunction.\newline}}
\begin{tabular}{c|ccc|cc|c}
\hline\hline\rule{0pt}{2.5ex}
&\multicolumn{3}{c|}{Non-relativistic}&\multicolumn{2}{c|}{Relativistic}&\\
&$3q$&$3q+5q$ direct&$3q+5q$ dir.+ exch.&$3q$&$3q+5q$ direct&Exp.
value\\\hline\rule{0pt}{3ex}
$g^{(3)}_A$&5/3&1.359&1.360&1.435&1.241&1.257$\pm$0.003\\
$g^{(8)}_A$&$1/\sqrt{3}$&0.499&0.500&0.497&0.444&0.34$\pm$0.02\\
$g^{(0)}_A$&1&0.900&0.901&0.861&0.787&0.31$\pm$0.07\\
$\Delta u$&4/3&1.123&1.125&1.148&1.011&0.83$\pm$0.03\\
$\Delta d$&-1/3&-0.236&-0.235&-0.287&-0.230&-0.43$\pm$0.043\\
$\Delta s$&0&0.012&0.012&0&0.006&-0.10$\pm$0.03\\
$\uN^{(5)}/\uN^{(3)}$&--&0.536&0.550&--&0.289&--\\
\hline\hline
\end{tabular}\label{Nucleon results}\end{center}
\end{table}
Although the 5-quark contributions improve the too simplistic
3-quark view, one can see that the direct contributions are dominant
while the exchange ones are clearly negligible. This is partly due
to the small values of the integrals (\ref{Exchange values}) which
are phase-space suppressed compared to (\ref{Direct values}). One
can also notice that relativistic corrections have a non-negligible
impact on the observables (the relativistic correction to the
3-quark component of the axial charges amounts to a multiplication
of the non-relativistic values by a factor of 0.861) and then
conclude that the non-relativistic approximation is too crude. Since
non-relativistic exchange contributions change the observable so
little we haven't computed their relativistic corrections.

We have fairly well reproduced $g^{(3)}_A=1.241$ while the
experimental value is 1.257$\pm$0.003. However the computed axial
charges $g^{(8)}_A$ and $g^{(0)}_A$ are not satisfactory (0.444 and
0.787 against 0.34$\pm$0.02 and 0.31$\pm$0.07). Only additional
quark-antiquark pairs contribute to $\Delta s$. Unfortunately the
effect of one pair in our computation is in the wrong direction
since the contribution is \emph{positive}. On the top of that
relativistic effects and addition of a pair reduce the
non-relativistic 3-quark amplitude of $\Delta d$ instead of
increasing it. In order to preserve $g^{(3)}_A$, one should explain
the shift of 0.2 between experimental and computed values for
$\Delta u$ and $\Delta d$.

The axial charge of the $\Theta^+\to K^+n$ transition allows one to
roughly estimate the $\Theta^+$ width. If we assume the approximate
$SU(3)$ chiral symmetry one can obtain the $\Theta\to KN$
pseudoscalar coupling from the generalized Goldberger-Treiman
relation
\begin{equation}
g_{\Theta KN}=\frac{g_A(\Theta\to KN)(M_\Theta+M_N)}{2F_K}
\end{equation}
where we use $M_\Theta=1530$ MeV, $M_N=940$ MeV and
$F_K=1.2F_\pi=112$ MeV. Once this transition pseudoscalar constant
is known one can evaluate the $\Theta^+$ width from the general
expression for the $\frac{1}{2}^+$ hyperon decay \cite{Width}
\begin{equation}
\Gamma_\Theta=2\,\frac{g^2_{\Theta
KN}|\up|}{8\pi}\frac{(M_\Theta-M_N)^2-m_K^2}{M_\Theta^2}
\end{equation}
where
$|\up|=\sqrt{(M_\Theta^2-M_N^2-m_K^2)^2-4M_N^2m_K^2}/2M_\Theta=254$
MeV is the kaon momentum in the decay ($m_K=495$ MeV) and the factor
of 2 stands for the equal probability $K^+n$ and $K^0p$ decays. All
results for the $\Theta^+$ pentaquark are collected in Table
\ref{Theta results}.
\begin{table}[h!]\begin{center}\caption{\small{Results for the $\Theta^+$ pentaquark: axial charge of the $\Theta^+\to K^+n$ transition,
$\Theta\to KN$ pseudoscalar coupling and $\Theta^+$ width. First,
results in the non-relativistic approximation are given, then with
relativistic corrections to the discrete-level
wavefunction.\newline}}
\begin{tabular}{c|cc|c}
\hline\hline \rule{0pt}{2.5ex}&\multicolumn{2}{c|}{Non-relativistic}&\multicolumn{1}{c}{Relativistic}\\
&$3q+5q$ direct&$3q+5q$ dir.+ exch.&$3q+5q$ direct\\
\hline\rule{0pt}{3ex}
$g_A(\Theta\to KN)$&0.202&0.203&0.144\\
$g_{\Theta KN}$&2.230&2.242&1.592\\
$\Gamma_\Theta$ (MeV)&4.427&4.472&2.256\\ \hline\hline
\end{tabular}\label{Theta results}\end{center}
\end{table}
Such as in the nucleon case, the exchange contribution is
negligible. However, relativistic corrections to the discrete-level
wavefunction are not negligible (reduction of 30\% for the axial
coupling and of 50\% for the width). This can be expected from the
fact that the $\Theta^+$ width directly depends on the number of
$q\bar q$ pairs in ordinary baryons \cite{Original paper}. Indeed,
the axial transition from the $\Theta^+$ to a nucleon can only take
place between similar Fock components. This means that the 5-quark
component of the $\Theta^+$ can only be connected with the 5-quark
component of the nucleon. Since relativistic corrections reduce the
5- to 3-quark normalization of the nucleon, so is the $\Theta^+$
width.

\section{Conclusion}

The Chiral Quark Soliton Model \cite{Profile function} provides a
relativistic description of the light baryons with an indefinite
number of $q\bar q$ pairs. Using this model, Diakonov and Petrov
\cite{Original paper} have presented a technique allowing one to
write down explicitly the 3-, 5-, 7-, \dots quark wavefunctions of
the octet, decuplet and antidecuplet. It is important that the
$q\bar q$ pair in the 5-quark component of any baryon is added in
the form of a chiral field, which costs little energy. That is why
the 5-quark component of the nucleon turns out to be substantial and
why the exotic $\Theta^+$ baryon is expected to be light.

For self-consistency this technique has been reminded and then used
in the present paper. It is really powerful and with sufficiently
patience one can write any Fock component of any baryon and compute
lots of matrix elements. Diakonov and Petrov have estimated the
normalization of the 5-quark component of the nucleon as about 50\%
of the 3-quark component, meaning that about 1/3 of the time the
nucleon is \emph{made} of five quarks. They have also showed that
the 5-quark component in the nucleon moves its axial charge
$g_A(p\to\pi^+n)$ from the naive non-relativistic value 5/3 much
closer to the experimental value. They have estimated the $\Theta^+$
width as being $\sim 4$ MeV thanks to the axial constant for the
$\Theta\to KN$ transition and showed that it is proportional to the
number of $q\bar q$ pairs in ordinary baryons. Assuming $SU(3)$
symmetry, the $\Theta^+$ width is additionally suppressed by the
$SU(3)$ Clebsch-Gordan factors. Therefore, the $\Theta^+$ width of a
few MeV appears naturally in the Chiral Quark Soliton Model without
any parameter fixing.

However, these estimations are rather crude since several
approximations were used (the first-order perturbation theory in
$1-\epsilon$ where $\epsilon=E_\textrm{lev}/M\sim 0.58$): the lower
component of the valence wavefunction $j(r)$ was ignored as well as
the distortion of the valence wavefunction by the sea, an
approximate expression for the pair wavefunction was used, the 7-,
9-, \ldots quark components were neglected and exchange
contributions to the 5-quark component were disregarded. It is
difficult to evaluate the errors of these approximations.
Unfortunately, the uncertainty associated with this non-relativistic
approximation is expected to be large since the expansion parameter
$1-\epsilon=0.42$ is poor. Another sign saying that the nucleon is a
relativistic system comes from the 50\% ratio of the 5-quark to the
3-quark normalization. It was also expected that exchange
contributions reduce further the $\Theta^+$ width and that is what
actually motivated the present work.

We have improved the technique by taking into account on the one
hand the 5-quark exchange contributions and on the other hand
relativistic corrections to the discrete-level wavefunction. Due to
the relative sign of their contributions, the 5-quark exchange
diagrams were expected to be a main source of error. In fact it
turns out that they are completely negligible, a fact partly due to
the phase-space suppression of the integrals. The other main source
of uncertainty was the relativistic approximation. This time, as
expected from the hints that the nucleon is a genuine relativistic
system, the relativistic corrections have a non-negligible impact on
observables. Especially, they reduce the 5- to 3-quark normalization
of the nucleon to 30\% instead of 50\%. This has the direct effect
to reduce also the $\Theta^+$ width which has now been estimated to
$\sim 2$ MeV. We have also computed all nucleon axial charges. Even
if we find $g_A^{(3)}=1.241$, $g_A^{(8)}$ and $g_A^{(0)}$ are not
satisfactory, especially the latter (0.444 and 0.787 against
0.34$\pm$0.02 and 0.31$\pm$0.07). The $\Delta s$ then obtained is
small and positive (0,006 against -0.10$\pm$0.03) while $\Delta u$
and $\Delta d$ are both 0.2 higher than the experimental values
(1.017 and -0.230 against 0.83$\pm$0.03 and -0.43$\pm$0.043).

The distortion of the valence level due to the sea has been
neglected and has probably another non-negligible effect on the
observables. The 7-, 9-, \ldots quark Fock components are not
believed to have a strong impact. Nevertheless it is rather
difficult to estimate the impact unless an explicit computation is
done.

The formalism has a broad field of applications, apart from exotic
baryons. One can indeed compute any type of transition amplitudes
between various Fock components of baryons, including the
relativistic effects, the effects of the $SU(3)$ symmetry violation,
the mixing of multiplets and so on. One can then in principle study
various vector and axial charges, the magnetic moments and magnetic
transitions, derive parton distributions thanks to this technique.

\subsection*{Acknowledgements}

The author is grateful to RUB TP2 for its kind hospitality, to D.
Diakonov and V. Petrov for enlightening discussions, explanations
and help. M. Polyakov is also thanked for his careful reading and
comments. The author is also indebted to J. Cugnon whose absence
would not have permitted the present work to be done. This work has
been supported by the National Funds of Scientific Research,
Belgium.
\newpage
\appendix

\renewcommand{\theequation}{A\arabic{equation}}
\setcounter{equation}{0}

\section*{Appendix A: Group integrals}

We give in this appendix a list of group integrals over the Haar
measure of the $SU(N)$ group and normalized to unity $\int\ud R=1$
that are needed for the technique. Most of them are simply copied
from the Appendix B of \cite{Original paper}. For the sake of
completeness we have also added the group integral that allows one
to derive the 5-quark component of the decuplet baryons.

For any $SU(N)$ group one has
\begin{equation}
\int\ud R\,R^f_i=0,\qquad \int\ud R\,R^{\dag i}_f=0,\qquad \int\ud
R\, R^f_iR^{\dag j}_g=\frac{1}{N}\,\delta^f_g\delta^j_i.
\end{equation}
For $N=2$, the following group integral is non-zero
\begin{equation}
\int\ud R\,R^f_iR^g_j=\frac{1}{2}\,\epsilon^{fg}\epsilon_{ij}
\end{equation}
while it is zero for $N>2$. The $SU(3)$ analog is
\begin{equation}
\int\ud R\,R^f_iR^g_jR^h_k=\frac{1}{6}\,\epsilon^{fgh}\epsilon_{ijk}
\end{equation}
which is on the contrary zero for $SU(2)$.

Here is the general method of finding integrals of several matrices
$R$, $R^\dag$. The result of an integration over the invariant
measure can be only invariant tensors which, for the $SU(N)$ group,
can be built solely from the Kronecker $\delta$ and Levi-Civita
$\epsilon$ tensors. One constructs the supposed tensor of a given
rank as a combination of $\delta$'s and $\epsilon$'s, satisfying the
symmetry relations following from the integral in question. The
indefinite coefficients in the combination are then found from
contracting both sides with various $\delta$'s and $\epsilon$'s and
thus by reducing the integral to a previously derived one.

For any $SU(N)$ group one has
\begin{equation}
\int\ud R\,R^{f_1}_{i_1}R^{\dag j_1}_{g_1}R^{f_2}_{i_2}R^{\dag
j_2}_{g_2}=\frac{1}{N^2-1}\left[\delta^{f_1}_{g_1}\delta^{f_2}_{g_2}\left(\delta^{j_1}_{i_1}\delta^{j_2}_{i_2}-\frac{1}{N}\,\delta^{j_2}_{i_1}\delta^{j_1}_{i_2}\right)
+\delta^{f_1}_{g_2}\delta^{f_2}_{g_1}\left(\delta^{j_2}_{i_1}\delta^{j_1}_{i_2}-\frac{1}{N}\,\delta^{j_1}_{i_1}\delta^{j_2}_{i_2}\right)\right].
\end{equation}

In $SU(2)$ there is an identity
\begin{equation}
\delta^j_{j_1}\epsilon_{j_2j_3}+\delta^j_{j_2}\epsilon_{j_3j_1}+\delta^j_{j_3}\epsilon_{j_1j_2}=0,\label{Identity}
\end{equation}
using which one finds that the following integral is non-zero
\begin{equation}
\int\ud R\,R^{f_1}_{j_1}R^{f_2}_{j_2}R^{f_3}_{j_3}R^{\dag
j}_{g}=\frac{1}{6}\left(\delta^{f_1}_{g}\delta^j_{j_1}\epsilon^{f_2f_3}\epsilon_{j_2j_3}+\delta^{f_2}_{g}\delta^j_{j_2}\epsilon^{f_3f_1}\epsilon_{j_3j_1}
+\delta^{f_3}_{g}\delta^j_{j_3}\epsilon^{f_1f_2}\epsilon_{j_1j_2}\right).
\end{equation}
For $N>2$ this integral is zero. The analog of the identity
(\ref{Identity}) in $SU(3)$ is
\begin{equation}
\delta^j_{j_1}\epsilon_{j_2j_3j_4}-\delta^j_{j_2}\epsilon_{j_3j_4j_1}+\delta^j_{j_3}\epsilon_{j_4j_1j_2}-\delta^j_{j_4}\epsilon_{j_1j_2j_3}=0,
\end{equation}
which gives the group integral involved when an octet baryon is
projected onto three quarks
\begin{eqnarray}
&&\int\ud
R\,R^{f_1}_{j_1}R^{f_2}_{j_2}R^{f_3}_{j_3}R^{f_4}_{j_4}R^{\dag
j}_{g}\nonumber\\
&=&\frac{1}{24}\left(\delta^{f_1}_{g}\delta^j_{j_1}\epsilon^{f_2f_3f_4}\epsilon_{j_2j_3j_4}+\delta^{f_2}_{g}\delta^j_{j_2}\epsilon^{f_3f_4f_1}\epsilon_{j_3j_4j_1}
+\delta^{f_3}_{g}\delta^j_{j_3}\epsilon^{f_4f_1f_2}\epsilon_{j_4j_1j_2}+\delta^{f_4}_{g}\delta^j_{j_4}\epsilon^{f_1f_2f_3}\epsilon_{j_1j_2j_3}\right).
\end{eqnarray}

To evaluate the $SU(3)$ average of six matrices, one needs the
identities
\begin{eqnarray}
\epsilon_{i_1j_2j_3}\epsilon_{j_1i_2i_3}+\epsilon_{i_2j_2j_3}\epsilon_{i_1j_1i_3}+\epsilon_{i_3j_2j_3}\epsilon_{i_1i_2j_1}&=&\nonumber\\
\epsilon_{j_1i_1j_3}\epsilon_{j_2i_2i_3}+\epsilon_{j_1i_2j_3}\epsilon_{i_1j_2i_3}+\epsilon_{j_1i_3j_3}\epsilon_{i_1i_2j_2}&=&\nonumber\\
\epsilon_{j_1j_2i_1}\epsilon_{j_3i_2i_3}+\epsilon_{j_1j_2i_2}\epsilon_{i_1j_3i_3}+\epsilon_{j_1j_2i_3}\epsilon_{i_1i_2j_3}&=&0.
\end{eqnarray}
One gets then the group integral involved when an antidecuplet
baryon is projected onto three quarks
\begin{eqnarray}
&&\int\ud
R\,R^{f_1}_{j_1}R^{f_2}_{j_2}R^{f_3}_{j_3}R^{h_1}_{i_1}R^{h_2}_{i_2}R^{h_3}_{i_3}=\frac{1}{72}\left(\epsilon^{f_1f_2f_3}\epsilon^{h_1h_2h_3}\epsilon_{j_1j_2j_3}\epsilon_{i_1i_2i_3}\right.\nonumber\\
&+&\epsilon^{h_1f_2f_3}\epsilon^{f_1h_2h_3}\epsilon_{i_1j_2j_3}\epsilon_{j_1i_2i_3}+\epsilon^{h_2f_2f_3}\epsilon^{h_1f_1h_3}\epsilon_{i_2j_2j_3}\epsilon_{i_1j_1i_3}+\epsilon^{h_3f_2f_3}\epsilon^{h_1h_2f_1}\epsilon_{i_3j_2j_3}\epsilon_{i_1i_2j_1}\nonumber\\
&+&\epsilon^{f_1h_1f_3}\epsilon^{f_2h_2h_3}\epsilon_{j_1i_1j_3}\epsilon_{j_2i_2i_3}+\epsilon^{f_1h_2f_3}\epsilon^{h_1f_2h_3}\epsilon_{j_1i_2j_3}\epsilon_{i_1j_2i_3}+\epsilon^{f_1h_3f_3}\epsilon^{h_1h_2f_2}\epsilon_{j_1i_3j_3}\epsilon_{i_1i_2j_2}\nonumber\\
&+&\epsilon^{f_1f_2h_1}\epsilon^{f_3h_2h_3}\epsilon_{j_1j_2i_1}\epsilon_{j_3i_2i_3}+\epsilon^{f_1f_2h_2}\epsilon^{h_1f_3h_3}\epsilon_{j_1j_2i_2}\epsilon_{i_1j_3i_3}+\epsilon^{f_1f_2h_3}\epsilon^{h_1h_2f_3}\epsilon_{j_1j_2i_3}\epsilon_{i_1i_2j_3}\big).\label{Three
quarks antidecuplet}
\end{eqnarray}

The result for the next integral is rather lengthy. We give it for
the general $SU(N)$. For abbreviation, we use the notation
\begin{equation}
\delta^{f_1}_a\delta^{f_2}_b\delta^{f_3}_c\delta^d_{j_1}\delta^e_{j_2}\delta^f_{j_3}
\equiv (abc)(def).\label{Notation integrals}
\end{equation}
One has the following group integral involved when a decuplet baryon
is projected onto three quarks
\begin{eqnarray}
&&\int\ud R\,R^{f_1}_{j_1}R^{f_2}_{j_2}R^{f_3}_{j_3}R_{h_1}^{\dag
i_1}R_{h_2}^{\dag i_2}R_{h_3}^{\dag
i_3}=\frac{1}{N(N^2-1)(N^2-4)}\nonumber\\
&\times&\{(N^2-2)\,[(123)(123)+(132)(132)+(321)(321)+(213)(213)+(312)(312)+(231)(231)]\nonumber\\
&-&N\,[(123)\,((132)+(321)+(213))+(132)\,((123)+(231)+(312))+(321)\,((312)+(123)+(231))\nonumber\\
&+&(213)\,((231)+(312)+(123))+(312)\,((213)+(132)+(321))+(231)\,((321)+(213)+(132))]\nonumber\\
&+&2\,[(123)\,((312)+(231))+(132)\,((213)+(321))+(321)\,((132)+(213))\nonumber\\
&+&(213)\,((321)+(132))+(312)\,((123)+(231))+(231)\,((312)+(123))]\}.
\end{eqnarray}
Apparently at $N=2$ something gets wrong. For $N=2$ there is a
formal identity following from the fact that one has for this
special case $\epsilon^{f_1f_2f_3}\epsilon_{h_1h_2h_3}=0$
\begin{equation}
(123)+(231)+(312)-(132)-(321)-(213)=0.
\end{equation}
Consequently, for $SU(2)$ one obtains a shorter expression
\begin{eqnarray}
&&\int\ud R\,R^{f_1}_{j_1}R^{f_2}_{j_2}R^{f_3}_{j_3}R_{h_1}^{\dag
i_1}R_{h_2}^{\dag i_2}R_{h_3}^{\dag
i_3}\nonumber\\
&=&\frac{1}{6}\,\{[(123)(123)+(132)(132)+(321)(321)+(213)(213)+(312)(312)+(231)(231)]\nonumber\\
&-&\frac{1}{4}\,[(123)\,((132)+(321)+(213))+(132)\,((123)+(231)+(312))+(321)\,((312)+(123)+(231))\nonumber\\
&+&(213)\,((231)+(312)+(123))+(312)\,((213)+(132)+(321))+(231)\,((321)+(213)+(132))]\}
\end{eqnarray}

If one is interested in the presence of an additional
quark-antiquark pair in an octet baryon, one has to use the group
integral
\begin{eqnarray}
&&\int\ud
R\,R^{f_1}_{j_1}R^{f_2}_{j_2}R^{f_3}_{j_3}\left(R^{f_4}_{j_4}R^{\dag
j_5}_{f_5}\right)R^h_3R^{\dag k}_g\nonumber\\
&=&\frac{1}{360}\left\{\epsilon^{f_1f_2h}\epsilon_{j_1j_2}\left[\delta^{f_3}_g\delta^{f_4}_{f_5}\left(4\delta^{j_5}_{j_4}\delta^k_{j_3}-\delta^{j_5}_{j_3}\delta^k_{j_4}\right)+\delta^{f_4}_g\delta^{f_3}_{f_5}\left(4\delta^{j_5}_{j_3}\delta^k_{j_4}-\delta^{j_5}_{j_4}\delta^k_{j_3}\right)\right]\right.\nonumber\\
&+&\epsilon^{f_1f_3h}\epsilon_{j_1j_3}\left[\delta^{f_2}_g\delta^{f_4}_{f_5}\left(4\delta^{j_5}_{j_4}\delta^k_{j_2}-\delta^{j_5}_{j_2}\delta^k_{j_4}\right)+\delta^{f_4}_g\delta^{f_2}_{f_5}\left(4\delta^{j_5}_{j_2}\delta^k_{j_4}-\delta^{j_5}_{j_4}\delta^k_{j_2}\right)\right]\nonumber\\
&+&\epsilon^{f_1f_4h}\epsilon_{j_1j_4}\left[\delta^{f_2}_g\delta^{f_3}_{f_5}\left(4\delta^{j_5}_{j_3}\delta^k_{j_2}-\delta^{j_5}_{j_2}\delta^k_{j_3}\right)+\delta^{f_3}_g\delta^{f_2}_{f_5}\left(4\delta^{j_5}_{j_2}\delta^k_{j_3}-\delta^{j_5}_{j_3}\delta^k_{j_2}\right)\right]\nonumber\\
&+&\epsilon^{f_2f_3h}\epsilon_{j_2j_3}\left[\delta^{f_1}_g\delta^{f_4}_{f_5}\left(4\delta^{j_5}_{j_4}\delta^k_{j_1}-\delta^{j_5}_{j_1}\delta^k_{j_4}\right)+\delta^{f_4}_g\delta^{f_1}_{f_5}\left(4\delta^{j_5}_{j_1}\delta^k_{j_4}-\delta^{j_5}_{j_4}\delta^k_{j_1}\right)\right]\nonumber\\
&+&\epsilon^{f_2f_4h}\epsilon_{j_2j_4}\left[\delta^{f_1}_g\delta^{f_3}_{f_5}\left(4\delta^{j_5}_{j_3}\delta^k_{j_1}-\delta^{j_5}_{j_1}\delta^k_{j_3}\right)+\delta^{f_3}_g\delta^{f_1}_{f_5}\left(4\delta^{j_5}_{j_1}\delta^k_{j_3}-\delta^{j_5}_{j_3}\delta^k_{j_1}\right)\right]\nonumber\\
&+&\epsilon^{f_3f_4h}\epsilon_{j_3j_4}\left[\delta^{f_1}_g\delta^{f_2}_{f_5}\left(4\delta^{j_5}_{j_2}\delta^k_{j_1}-\delta^{j_5}_{j_1}\delta^k_{j_2}\right)+\delta^{f_2}_g\delta^{f_1}_{f_5}\left(4\delta^{j_5}_{j_1}\delta^k_{j_2}-\delta^{j_5}_{j_2}\delta^k_{j_1}\right)\right]\nonumber
\end{eqnarray}
\begin{eqnarray}
&+&\epsilon^{f_1f_2f_3}\epsilon_{j_1j_2j_3}\left[\delta^h_g\delta^{f_4}_{f_5}\left(4\delta^{j_5}_{j_4}\delta^k_3-\delta^{j_5}_3\delta^k_{j_4}\right)+\delta^{f_4}_g\delta^h_{f_5}\left(4\delta^{j_5}_3\delta^k_{j_4}-\delta^{j_5}_{j_4}\delta^k_3\right)\right]\nonumber\\
&+&\epsilon^{f_2f_3f_4}\epsilon_{j_2j_3j_4}\left[\delta^h_g\delta^{f_1}_{f_5}\left(4\delta^{j_5}_{j_1}\delta^k_3-\delta^{j_5}_3\delta^k_{j_1}\right)+\delta^{f_1}_g\delta^h_{f_5}\left(4\delta^{j_5}_3\delta^k_{j_1}-\delta^{j_5}_{j_1}\delta^k_3\right)\right]\nonumber\\
&+&\epsilon^{f_3f_4f_1}\epsilon_{j_3j_4j_1}\left[\delta^h_g\delta^{f_2}_{f_5}\left(4\delta^{j_5}_{j_2}\delta^k_3-\delta^{j_5}_3\delta^k_{j_2}\right)+\delta^{f_2}_g\delta^h_{f_5}\left(4\delta^{j_5}_3\delta^k_{j_2}-\delta^{j_5}_{j_2}\delta^k_3\right)\right]\nonumber\\
&+&\epsilon^{f_4f_1f_2}\epsilon_{j_4j_1j_2}\left[\delta^h_g\delta^{f_3}_{f_5}\left(4\delta^{j_5}_{j_3}\delta^k_3-\delta^{j_5}_3\delta^k_{j_3}\right)+\delta^{f_3}_g\delta^h_{f_5}\left(4\delta^{j_5}_3\delta^k_{j_3}-\delta^{j_5}_{j_3}\delta^k_3\right)\right]\Big\}.
\end{eqnarray}

For finding the quark structure of the antidecuplet, the following
group integrals are relevant. The conjugate rotational wavefunction
of the antidecuplet is
\begin{equation}
A_k^{*\{h_1h_2h_3\}}(R)=\frac{1}{3}\left(R^{h_1}_3R^{h_2}_3R^{h_3}_k+R^{h_2}_3R^{h_3}_3R^{h_1}_k+R^{h_3}_3R^{h_1}_3R^{h_2}_k\right).\label{Tensor
pentaquarks}
\end{equation}
Projecting it on three quarks and using eq. (\ref{Three quarks
antidecuplet}) one gets an identical zero because all terms in
(\ref{Three quarks antidecuplet}) are antisymmetric in a pair of
flavor indices while the tensor (\ref{Tensor pentaquarks}) is
symmetric. It reflects the fact that one cannot build an
antidecuplet from three quarks
\begin{equation}
\int\ud
R\,R^{f_1}_{j_1}R^{f_2}_{j_2}R^{f_3}_{j_3}A_k^{*\{h_1h_2h_3\}}(R)=0.
\end{equation}
However, a similar group integral with an additional quark-antiquark
pair is non-zero
\begin{eqnarray}
&&\int\ud
R\,R^{f_1}_{j_1}R^{f_2}_{j_2}R^{f_3}_{j_3}\left(R^{f_4}_{j_4}R^{\dag
j_5}_{f_5}\right)A_k^{*\{h_1h_2h_3\}}(R)=\frac{1}{1080}\nonumber\\
&\times&\left\{\left(\delta^{j_5}_k\epsilon_{j_1j_2}\epsilon_{j_3j_4}+\delta^{j_5}_3\epsilon_{j_1j_2k}\epsilon_{j_3j_4}+\delta^{j_5}_3\epsilon_{j_1j_2}\epsilon_{j_3j_4k}\right)\left[\delta^{h_3}_{f_5}\left(\epsilon^{f_1f_2h_1}\epsilon^{f_3f_4h_2}+\epsilon^{f_1f_2h_2}\epsilon^{f_3f_4h_1}\right)\right.\right.\nonumber\\
&+&\delta^{h_1}_{f_5}\left(\epsilon^{f_1f_2h_3}\epsilon^{f_3f_4h_2}+\epsilon^{f_1f_2h_2}\epsilon^{f_3f_4h_3}\right)+\delta^{h_2}_{f_5}\left(\epsilon^{f_1f_2h_1}\epsilon^{f_3f_4h_3}+\epsilon^{f_1f_2h_3}\epsilon^{f_3f_4h_1}\right)\Big]\nonumber\\
&+&\left(\delta^{j_5}_k\epsilon_{j_2j_3}\epsilon_{j_4j_1}+\delta^{j_5}_3\epsilon_{j_2j_3k}\epsilon_{j_4j_1}+\delta^{j_5}_3\epsilon_{j_2j_3}\epsilon_{j_4j_1k}\right)\left[\delta^{h_3}_{f_5}\left(\epsilon^{f_2f_3h_1}\epsilon^{f_4f_1h_2}+\epsilon^{f_2f_3h_2}\epsilon^{f_4f_1h_1}\right)\right.\nonumber\\
&+&\delta^{h_1}_{f_5}\left(\epsilon^{f_2f_3h_3}\epsilon^{f_4f_1h_2}+\epsilon^{f_2f_3h_2}\epsilon^{f_4f_1h_3}\right)+\delta^{h_2}_{f_5}\left(\epsilon^{f_2f_3h_1}\epsilon^{f_4f_1h_3}+\epsilon^{f_2f_3h_3}\epsilon^{f_4f_1h_1}\right)\Big]\nonumber\\
&+&\left(\delta^{j_5}_k\epsilon_{j_1j_3}\epsilon_{j_2j_4}+\delta^{j_5}_3\epsilon_{j_1j_3k}\epsilon_{j_2j_4}+\delta^{j_5}_3\epsilon_{j_1j_3}\epsilon_{j_2j_4k}\right)\left[\delta^{h_3}_{f_5}\left(\epsilon^{f_1f_3h_1}\epsilon^{f_2f_4h_2}+\epsilon^{f_1f_3h_2}\epsilon^{f_2f_4h_1}\right)\right.\nonumber\\
&+&\delta^{h_1}_{f_5}\left(\epsilon^{f_1f_3h_3}\epsilon^{f_2f_4h_2}+\epsilon^{f_1f_3h_2}\epsilon^{f_2f_4h_3}\right)+\delta^{h_2}_{f_5}\left(\epsilon^{f_1f_3h_1}\epsilon^{f_2f_4h_3}+\epsilon^{f_1f_3h_3}\epsilon^{f_2f_4h_1}\right)\Big]\Big\}.
\end{eqnarray}

We complete this set of integrals by adding the projection of a
decuplet baryon onto three quarks and a quark-antiquark pair. The
result is rather lengthy. We introduce on the top of (\ref{Notation
integrals}) the following notation
\begin{eqnarray}
[abcd]&\equiv&(1234)(abcd)+(2341)(bcda)+(3412)(cdab)+(4123)(dabc)\nonumber\\
&+&(2134)(bacd)+(1342)(acdb)+(3421)(cdba)+(4213)(dbac)\nonumber\\
&+&(3214)(cbad)+(2143)(badc)+(1432)(adcb)+(4321)(dcba)\nonumber\\
&+&(4231)(dbca)+(2314)(bcad)+(3142)(cadb)+(1423)(adbc)\nonumber\\
&+&(1324)(acbd)+(3241)(cbda)+(2413)(bdac)+(4132)(dacb)\nonumber\\
&+&(1243)(abdc)+(2431)(bdca)+(4312)(dcab)+(3124)(cabd).
\end{eqnarray}
We then obtain
\begin{eqnarray}
&&\int\ud
R\,R^{f_1}_{j_1}R^{f_2}_{j_2}R^{f_3}_{j_3}R^{f_4}_{j_4}R_{h_1}^{\dag
i_1}R_{h_2}^{\dag i_2}R_{h_3}^{\dag i_3}R_{h_4}^{\dag
i_4}=\frac{1}{N^2(N^2-1)(N^2-4)(N^2-9)}\nonumber\\
&\times&\{(N^4-8N^2+6)[1234]-5N\left([2341]+[4123]+[3421]+[4312]+[3142]+[2413]\right)\nonumber\\
&+&(N^2+6)\left([3412]+[2143]+[4321]\right)-N(N^2-4)\left([2134]+[3214]+[1432]+[1324]+[1243]+[4231]\right)\nonumber\\
&+&(2N^2-3)\left([1342]+[4213]+[3241]+[2314]+[3124]+[4132]+[2431]+[1423]\right)\}.
\end{eqnarray}
There seems to be a problem when $N=2$ or $N=3$. There are however
formal identities that have to be taken into account leading to
shorter and well defined expressions. For $N=3$ we have
$\epsilon^{f_1f_2f_3f_4}\epsilon_{h_1h_2h_3h_4}=0$
\begin{eqnarray}
&&(1234)-(2341)+(3412)-(4123)+(2314)-(3142)+(1423)-(4231)\nonumber\\
&+&(3124)-(1243)+(2431)-(4312)-(1324)+(3241)-(2413)+(4132)\nonumber\\
&-&(3214)+(2143)-(1432)+(4321)-(2134)+(1342)-(3421)+(4213)=0.
\end{eqnarray}
Consequently, for $SU(3)$ we obtain the shorter expression
\begin{eqnarray}
&&\int\ud
R\,R^{f_1}_{j_1}R^{f_2}_{j_2}R^{f_3}_{j_3}R^{f_4}_{j_4}R_{h_1}^{\dag
i_1}R_{h_2}^{\dag i_2}R_{h_3}^{\dag i_3}R_{h_4}^{\dag
i_4}=\frac{1}{2160}\nonumber\\
&\times&\{48[1234]+7\left([2341]+[4123]+[3421]+[4312]+[3142]+[2413]\right)\nonumber\\
&-&6\left([3412]+[2143]+[4321]\right)+11\left([2134]+[3214]+[1432]+[1324]+[1243]+[4231]\right)\}.
\end{eqnarray}
For $N=2$ on the one hand we have
$\delta^{f_1}_a\epsilon^{f_2f_3f_4}\epsilon_{bcd}=\delta^{f_2}_b\epsilon^{f_3f_4f_1}\epsilon_{cda}=
\delta^{f_3}_c\epsilon^{f_4f_1f_2}\epsilon_{dab}=\delta^{f_4}_d\epsilon^{f_1f_2f_3}\epsilon_{abc}=0$
\begin{eqnarray}
&&(abcd)+(acdb)+(adbc)-(acbd)-(abdc)-(adcb)=0,\\
&&(abcd)+(cbda)+(dbac)-(cbad)-(abdc)-(dbca)=0,\\
&&(abcd)+(bdca)+(dacb)-(bacd)-(adcb)-(dbca)=0,\\
&&(abcd)+(bcad)+(cabd)-(bacd)-(acbd)-(cbad)=0.
\end{eqnarray}
On the other hand for $N=2$ we have
$\epsilon^{f_1f_2f_3f_4}\epsilon_{abk}\epsilon_{cdl}\epsilon^{kl}=\epsilon^{f_1f_3f_2f_4}\epsilon_{ack}\epsilon_{bdl}\epsilon^{kl}=\epsilon^{f_1f_4f_2f_3}\epsilon_{adk}\epsilon_{bcl}\epsilon^{kl}=0$
\begin{eqnarray}
&&(abcd)-(bacd)+(badc)-(abdc)-(cdab)+(cdba)-(dcba)+(dcab)=0,\\
&&(abcd)-(cbad)+(cdab)-(adcb)-(badc)+(dabc)-(dcba)+(bcda)=0,\\
&&(abcd)-(dbca)+(dcba)-(acbd)-(badc)+(cadb)-(cdab)+(bdac)=0.
\end{eqnarray}
Consequently, for $SU(2)$ we obtain the shorter expression
\begin{eqnarray}
&&\int\ud
R\,R^{f_1}_{j_1}R^{f_2}_{j_2}R^{f_3}_{j_3}R^{f_4}_{j_4}R_{h_1}^{\dag
i_1}R_{h_2}^{\dag i_2}R_{h_3}^{\dag i_3}R_{h_4}^{\dag
i_4}=\frac{-1}{240}\nonumber\\
&\times&\{16[1234]+11\left([2341]+[4123]+[3421]+[4312]+[3142]+[2413]\right)\nonumber\\
&-&12\left([3412]+[2143]+[4321]\right)-8\left([2134]+[3214]+[1432]+[1324]+[1243]+[4231]\right)\}.
\end{eqnarray}

\renewcommand{\theequation}{B\arabic{equation}}
\setcounter{equation}{0}

\section*{Appendix B: General tools for the $n$-quark Fock
component}

In this appendix we will give general remarks and ``tricks'' that
help to derive easily the contributions of \emph{any} Fock
component. We will show that schematic diagrams drawed by Diakonov
and Petrov \cite{Original paper} are a key tool that allows one to
rapidly give the sign, the spin-flavor structure, the number of
equivalent annihilation-creation operator contractions and the
factor coming from color contractions for any such diagram. We first
give the rules and then apply them to the 7-quark Fock component.

\begin{enumerate}
\item
First remember that dark gray rectangles of the diagrams stand for
the three valence quarks and light gray rectangles for
quark-antiquark pairs. Each line represents the color, flavor and
spin contractions
\begin{equation}
\delta^{\alpha_i}_{\alpha'_i}\delta^{f_i}_{f'_i}\delta^{\sigma_i}_{\sigma'_i}
\int\ud
z'_i\,\ud^2\up'_{i\perp}\delta(z_i-z'_i)\delta^{(2)}(\up_{i\perp}-\up'_{i\perp}).
\end{equation}
The reversed arrow stands for the antiquark.
\item
For any $n$-quark Fock component there are $(n+3)/2$ quark creation
operators and $(n-3)/2$ antiquark creation operators. The total
number of annihilation-creation operator contractions is then
\begin{equation}
\left(\frac{n+3}{2}\right)!\left(\frac{n-3}{2}\right)!
\end{equation}
This means that for the 3-quark component there are 6
annihilation-creation operator contractions and 24 for the 5-quark
component.
\item
The number of line crossings $N$ gives the sign of the
annihilation-creation operator contractions $(-1)^N$. Indeed, any
line crossing represents an anticommutation of operators.
\item
The color structure of the valence quarks is
$\epsilon^{\alpha_1\alpha_2\alpha_3}$ and for the quark-antiquark
pair it is $\delta^{\alpha_4}_{\alpha_5}$. So if one considers
color, the antiquark line and the quark line of the same pair can be
connected and then belong to the same circuit. The color factor is
at least 3! due to the contraction of both $\epsilon$'s with
possibly a minus sign. There is another factor of 3 for any circuit
that is not connected to the valence quarks.
\item
The valence quarks are equivalent which means that different
contractions of the same valence quarks are equivalent. Indeed any
sign coming from the crossings in rule 3 is compensated by the same
sign coming from the $\epsilon$ color contraction in rule 4. That is
the reason why one needs to draw only one diagram for the 3-quark
component.
\item
The quark-antiquark pairs are equivalent which means that any
vertical exchange of the light gray rectangles (quark and antiquark
lines stay fixed to the rectangles) does not produce a new type of
diagram. This appears only from the 7-quark component since one
needs at least two quark-antiquark pairs.
\end{enumerate}
So for the 5-quark component there are only two types of diagrams.
The direct one has no crossing and is thus positive while the
exchange one is negative due to one crossing. There are 6 equivalent
direct annihilation-creation contractions and the color factor is
$3!\cdot 3$ (there is an independent color circuit within the
quark-antiquark pair). There are 18 equivalent exchange
annihilation-creation contractions but the color factor is only 3!
since the pair lines belong to a valence circuit. This is exactly
what was said in subsection \ref{fivequarks diagrams}. Of course
there are $6+18=24$ annihilation-creation operator contractions for
the 5-quark component as stated by rule 2.

\begin{figure}[h]\begin{center}\includegraphics[width=15cm]{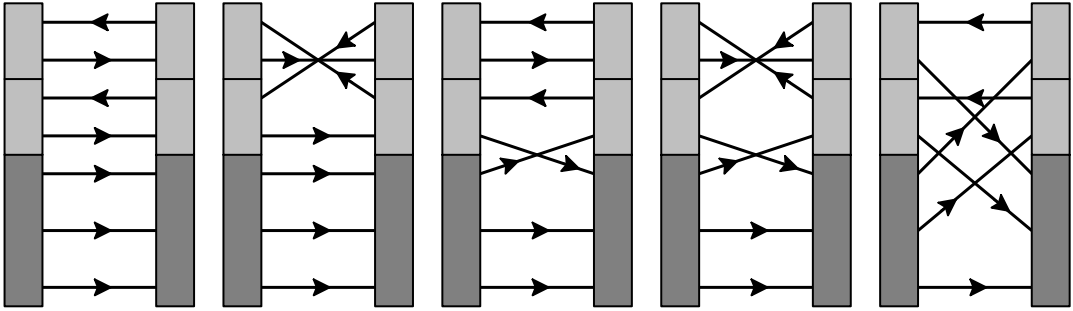}
\caption{\small{Schematic representation of the 7-quark
contributions to the
normalization.}}\label{Sevenquarks}\end{center}
\end{figure}

Let us now apply these rules to see what happens when one considers
the 7-quark Fock component. From rules 5 and 6 we obtain that there
are only five types of diagrams, see Fig \ref{Sevenquarks}.

Let us find the signs. These prototype diagrams have been chosen
such that color contractions do not affect the sign. The first
diagram is obviously positive (no crossing). The second one has
three crossings (they are degenerate in the drawing but it does not
change anything considering one or three crossings since the
important thing is that it is odd) and is thus negative. So is the
third one with its unique crossing. The fourth diagram has four
crossings and is thus positive. The last one has six crossings and
is thus also positive.
\begin{figure}[h]\begin{center}\begin{minipage}[c]{7cm}\begin{center}\includegraphics[width=3cm]{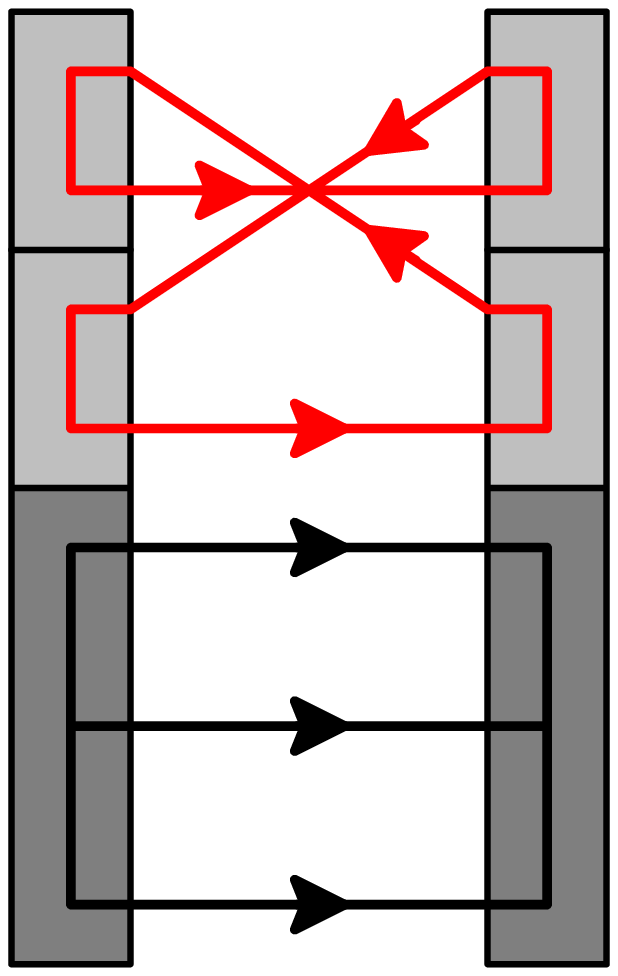}
\caption{\small{The color factor of this diagram is $3!\cdot 3$
since one has the valence circuit and an independent
circuit.\newline}}\label{Independent circuit}
\end{center}\end{minipage}\hspace{1cm}
\begin{minipage}[c]{7cm}\begin{center}\includegraphics[width=3cm]{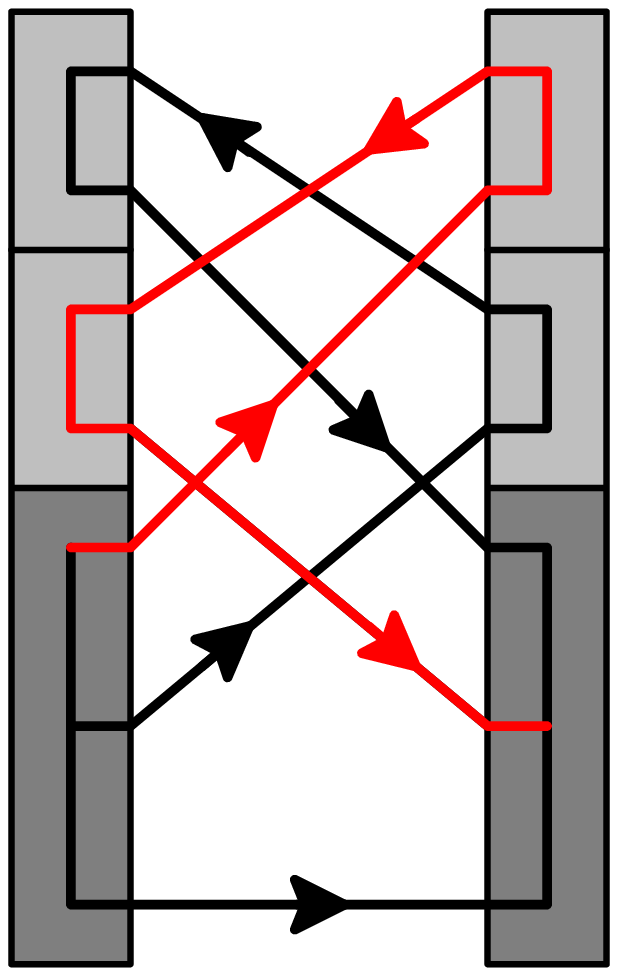}
\caption{\small{The color contractions in this diagram give a minus
factor because of interchange of two valence
quarks.\newline}}\label{Color sign}
\end{center}\end{minipage}\end{center}
\end{figure}

Following rule 2 there must be $5!2!=240$ contractions. Indeed,
there are 12 of the first and second types while there are 72 of the
other ones. Thus we have $2\cdot 12+3\cdot 72=240$ contractions as
expected.

The color factor of the first diagram is $3!\cdot 3\cdot 3=54$ since
there are two independent circuits. The color factor of the second
one is only $3!\cdot 3=18$ since there is only one independent
circuit as one can see on Fig. \ref{Independent circuit}. The third
diagram has also a unique independent circuit and thus a color
factor of $3!\cdot 3=18$. For the two last diagrams there are no
more independent circuit and have consequently a color factor of
$3!=6$.

We close this appendix by considering the diagram in Fig. \ref{Color
sign}. Since two valence quarks are exchanged, it must belong to the
fifth type of diagrams. There are seven crossings and thus a
negative sign while the fifth type of diagrams is positive. In fact,
for this particular diagram, the color contractions gives an
additional minus sign since the third quark on the left is
contracted with the second on the right
$\epsilon^{\alpha_1\alpha_2\alpha_3}\epsilon_{\alpha_1\alpha_3\alpha_2}=-6$.

\end{document}